\PassOptionsToPackage{table}{xcolor}

\documentclass[ 
reprint, 
nofootinbib, amsmath,amssymb, floatfix, aps, ]{revtex4-2}

\usepackage{graphicx}
\usepackage{dcolumn}
\usepackage{bm}

\usepackage[all]{xy}
\usepackage{xspace}
\usepackage{xfrac}
\usepackage{slashed}
\usepackage{hhline}
\usepackage{tabularx}

\usepackage{ulem}  
\usepackage{xcolor} 
\usepackage{float}

\usepackage{varioref}
\usepackage{hyperref}
\hypersetup{colorlinks=true,linkcolor=red!45!black,citecolor=green!40!black,urlcolor=blue!50!black,linktocpage=true}
\usepackage[capitalize]{cleveref}

\newcolumntype{L}{>{$}l<{$}}
\newcolumntype{C}{>{$}c<{$}}
\newcolumntype{R}{>{$}r<{$}}

\newcommand{\unitmatrix}{\openone}
\newcommand{\Lag}{\mathcal{L}}
\newcommand{\dbar}{\overline}
\newcommand{\hc}{\mathrm{h.c.}}
\newcommand{\Op}{\mathcal{O}}

\newcommand{\Lept}{\mathcal{L}}
\newcommand{\Bary}{\mathcal{B}}
\newcommand{\BminL}{{\left[B\! -\! L\right]}}

\newcommand{\BR}{\mathrm{BR}}
\newcommand{\BRV}{\mathrm{BR}_V}
\newcommand{\Ratio}{\mathrm{R}}
\newcommand{\RK}{\Ratio_K}
\newcommand{\RD}{\Ratio_D}

\newcommand{\RKs}{\Ratio_{K^*}}
\newcommand{\RKRKs}{\Ratio_{K^{(*)}}}
\newcommand{\RDRDs}{\Ratio_{D^{(*)}}}
\newcommand{\CR}{\mathrm{CR}}
\newcommand{\CRmueAu}{\CR(\mu\to e, \mathrm{Au})}

\newcommand{\KL}{K^0_L}
\newcommand{\KS}{K^0_S}
\newcommand{\KLS}{K^0_{L,S}}
\newcommand{\Bds}{B^0_{d,s}}

\newcommand{\MeV}{\;\mathrm{MeV}}
\newcommand{\TeV}{\;\mathrm{TeV}}
\newcommand{\GeV}{\;\mathrm{GeV}}

\newcommand{\flavio}{\texttt{flavio}\xspace}
\newcommand{\smelli}{\texttt{smelli}\xspace}
\newcommand{\wcxf}{\texttt{wcxf}\xspace}
\newcommand{\wilson}{\texttt{wilson}\xspace}

\newcommand{\WM}{\texttt{Wolfram Mathematica}\xspace}
\newcommand{\CKM}{V_\mathrm{CKM}}
\newcommand{\PMNS}{V_\mathrm{PMNS}}
\newcommand{\U}{{V}}
\newcommand{\UL}{{\U_{\!L}}}
\newcommand{\UR}{{\U_{\!R}}}
\newcommand{\ULp}{{\U_{\!L}'}}
\newcommand{\URp}{{\U_{\!R}'}}
\newcommand{\ULR}{{\U_{\!L,R}}}
\newcommand{\URL}{{\U_{\!R,L}}}

\newcommand{\Ul}[1]{(\UL)_{#1}}
\newcommand{\Ur}[1]{(\UR)_{#1}}
\newcommand{\UlR}[1]{(\ULR)_{#1}}
\newcommand{\UrL}[1]{(\URL)_{#1}}
\newcommand{\Ulep}{\U^\mathrm{I}}
\newcommand{\Uq}{\U^\mathrm{II}}
\newcommand{\Uqlep}{\U^\mathrm{III}}
\newcommand{\Unull}{\U^0}

\newcommand{\VdL}{{V^d_L}}

\newcommand{\VeL}{{V^e_L}}

\newcommand{\VdR}{{V^d_R}}

\newcommand{\VeR}{{V^e_R}}

\newcommand{\PL}{\mathbb{P}_{\kern -0.3mm L}}
\newcommand{\PR}{\mathbb{P}_{\kern -0.3mm R}}
\newcommand{\PLR}{\mathbb{P}_{\kern -0.3mm L,R}}

\usepackage[safe]{tipa}
\newcommand{\qubar}{{{\kern 0.2pt}\ipabar{$q$}{-1.pt}{0.35}{0.3}{0.1}{\kern 0.2pt}}}
\newcommand{\LeptonDoublet}{\ell}
\newcommand{\QuarkDoublet}{\qubar}
\newcommand{\QDoubletIndex}{{\scalebox{0.67}{\ipabar{$q$}{-1.pt}{0.35}{0.36}{0.16}}}}
\newcommand{\lep}{l}
\newcommand{\qu}{q} 
\newcommand{\ledq}{\LeptonDoublet e d \QDoubletIndex}
\newcommand{\nLext}{k_L}
\newcommand{\nRext}{k_R}
\newcommand{\nLRext}{k_{L,R}}
\newcommand{\Rhand}{{\scalebox{0.60}{$R$}}}
\newcommand{\Lhand}{{\scalebox{0.60}{$L$}}}
\newcommand{\uR}{{u_{\kern -0.4mm \Rhand}}}
\newcommand{\uL}{{u_{\kern -0.4mm \Lhand}}}
\newcommand{\dR}{{d_{\kern -0.35mm \Rhand}}}
\newcommand{\dL}{{d_{\kern -0.4mm \Lhand}}}
\newcommand{\eR}{{e_{\kern -0.4mm \Rhand}}}
\newcommand{\eL}{{e_{\kern -0.4mm \Lhand}}}
\newcommand{\nuR}{{\nu_{\kern -0.4mm \Rhand}}}
\newcommand{\nuL}{{\nu_{\kern -0.4mm \Lhand}}}
\newcommand{\LL}{{\LeptonDoublet_{\kern -0.4mm \Lhand}}}
\newcommand{\LR}{{\LeptonDoublet_{\kern -0.4mm \Rhand}}}
\newcommand{\LLR}{{\LeptonDoublet_{\kern -0.4mm \Lhand,\Rhand}}}
\newcommand{\eLR}{{e_{\kern -0.4mm \Lhand,\Rhand}}}

\newcommand{\QL}{{\QuarkDoublet_{\kern -0.3mm \Lhand}}}
\newcommand{\FQL}{{F_{\kern -0.4mm \Lhand}}}
\newcommand{\FQR}{{F_{\kern -0.4mm \Rhand}}}
\newcommand{\fuR}{{f^u_{\kern -0.4mm \Rhand}}}
\newcommand{\fdR}{{f^d_{\kern -0.4mm \Rhand}}}
\newcommand{\massbasis}[1]{\hat #1}
\newcommand{\uRmass}{{\massbasis u_{\kern -0.4mm \Rhand}}}
\newcommand{\uLmass}{{\massbasis u_{\kern -0.4mm \Lhand}}}
\newcommand{\dRmass}{{\massbasis d_{\kern -0.35mm \Rhand}}}
\newcommand{\dLmass}{{\massbasis d_{\kern -0.4mm \Lhand}}}
\newcommand{\eRmass}{{\massbasis e_{\kern -0.4mm \Rhand}}}
\newcommand{\eLmass}{{\massbasis e_{\kern -0.4mm \Lhand}}}
\newcommand{\nuRmass}{{\massbasis \nu_{\kern -0.4mm \Rhand}}}
\newcommand{\nuLmass}{{\massbasis \nu_{\kern -0.4mm \Lhand}}}
\newcommand{\LLmass}
	{{\massbasis \LeptonDoublet_{\kern -0.4mm \Lhand}}}
\newcommand{\QLmass}
	{{\massbasis \QuarkDoublet_{\kern -0.4mm \Lhand}}}
\newcommand{\downintbasis}[1]{\mathring{#1}}
\newcommand{\nuLint}{{\downintbasis\nu_{\kern -0.4mm \Lhand}}}

\newcommand{\begflmatrix}{\begin{pmatrix}}
\newcommand{\enflmatrix}{\end{pmatrix}}

\newcommand{\ala}{à~la\xspace}

\newcommand{\GPS}{G_\mathrm{422}}
\newcommand{\GQL}{G_\mathrm{421}}
\newcommand{\GBL}{G_{3121}}
\newcommand{\GSM}{G_\mathrm{SM}}
\newcommand{\Gvac}{G_\mathrm{vac}}

\newcommand{\SUL}{{SU(2)_L}}
\newcommand{\SUc}{{SU(3)_C}}
\newcommand{\SUC}{{SU(4)_C}}
\newcommand{\SUR}{{SU(2)_R}}

\newcommand{\BELLEII}{{\scshape{Belle~II}}\xspace}
\newcommand{\BABAR}{{\scshape{BaBar}}\xspace}
\newcommand{\LHCb}{{\scshape{LHC}}b\xspace}
\newcommand{\ATLAS}{{\scshape{Atlas}}\xspace}
\newcommand{\CMS}{{\scshape{CMS}}\xspace}
\newcommand{\collabName}[1]{{\scshape{#1}}\xspace}

\newcommand{\LF}{\text{LF}}
\newcommand{\LFU}{\text{LFU}}
\newcommand{\SM}{\mathrm{SM}}
\newcommand{\NP}{\mathrm{NP}}

\newcommand{\lflbracket}{(}
\newcommand{\rflbracket}{)}

\newcommand{\NN}{\mathcal{N}}
\newcommand{\sep}{\_\,}

\newcommand{\qL}{\qu_\Lhand}

\newcommand{\qLprBar}{\dbar{\qL'}}

\newcommand{\XXX}{\bullet}

\newcommand{\ZOprime}{\check Z'}

\begin{document}


\title{
\texorpdfstring{All possible first signals of gauge leptoquark \\in quark-lepton unification and beyond}{All possible first signals of gauge leptoquark in quark-lepton unification and beyond}}

\author{Hedvika Gedeonová}
 \email{gedeonova@ipnp.mff.cuni.cz}
\author{Matěj Hudec}%
 \email{hudec@ipnp.mff.cuni.cz (corresponding author)}
\affiliation{%
Institute of Particle and Nuclear Physics, Charles University\\
Prague, Czech Republic
}%


\date{October 2022 [v1];\, August 2023 [v2 = 
this version]}

\begin{abstract}
We study  possible current and future low-energy signals of the gauge leptoquark in quark-lepton $SU(4)$ unification \ala Pati-Salam. Taking fully into account the freedom in the generation mixing between quarks and leptons, we compile a catalogue of observables which currently form a border of the excluded part of the parameter space -- hot candidates for first signals of new physics. We also determine the sensitivity needed in order to inspect a currently allowed part of the parameter space for several other measurements which are not included in this catalogue.
We improve older similar works on this topic by taking into account more (and more recent) experimental measurements and by scanning the parameter space more densely.

Furthermore, we study in a similar manner the $SU(4)$ models with a small number of generations of extra leptons. We also discuss the minimal number of leptons needed in order to alleviate the contemporary discrepancies in the neutral-current $B$-meson decays.
  
\end{abstract}

\maketitle


\section{Introduction}\label{sec:intro}
The main goal of this phenomenological study is to list all possible smoking gun signals of the Pati-Salam leptoquark.

\subsection{Quark-lepton unification and gauge leptoquark}
\label{sec:intro:model}

Quark-lepton unification (QLU) \ala  Pati and Salam \cite{Pati:1974yy,Pati:1983zp} is an old idea motivated by the equal number of lepton and quark families and their similar electroweak behaviour.
Technically, QLU is based on 
extending the QCD gauge factor $\SUc$ to
$\SUC$ and accommodating the quarks and leptons in common 4-dimensional  representations:
\begin{align}
&\begin{pmatrix} \QL\\ \LL \end{pmatrix}\,,
&
&\begin{pmatrix} \uR\\ \nuR \end{pmatrix}\,,
&
&\begin{pmatrix} \dR\\ \eR \end{pmatrix}\,.
\label{fermionsIn4plets}
\end{align}

The most characteristic prediction of QLU is the existence of a gauge leptoquark (LQ) $U_1$ transforming as $(3,1,+\sfrac{2}{3})$ with respect to the Standard Model (SM) gauge group $\GSM = \SUc\times \SUL \times U(1)_Y$. The LQ has the following interactions with the fermions from \cref{fermionsIn4plets}:
\begin{align}
\Lag_\mathrm{int}&= 
\frac{g_4}{\sqrt 2}\left(
	 \dbar{\QLmass}_{i} \gamma_\mu \UL \LLmass^i
	+\dbar{\dRmass} \gamma_\mu \UR \eRmass
	+\dbar{\uRmass} \gamma_\mu \URp \nuRmass
\right) {U_1}^\mu 
\nonumber \\&+ \hc      
\label{Lag-U1}
\end{align}
Here $i\in\{1,2\}$ is an $\SUL$ index; $\dRmass, \uRmass,\eRmass, \nuRmass$ denote the family triplets of the same-charge fermions in the mass basis, e.g. 
$\dbar\dRmass = (\dbar{d_\Rhand}, \dbar{s^{\phantom{*}}_\Rhand}, \dbar{b_\Rhand})$, and similarly $\QLmass$ and $\LLmass$ are in the mass basis of their $T^3_L=-\sfrac12$ components. The $3\times 3$ flavour matrices $\UL,\UR,\URp$ and the LQ mass $m_{U_1}$ are free parameters of the theory.
QLU fixes the $g_4$ coupling at the scale of $\SUC$ breaking and restricts $\UL, \UR, \URp$ to unitary patterns, i.e.

\begin{subequations}
\begin{align}
g_4(m_{U_1})&=g_3(m_{U_1}), \\
\UL,\UR, \URp &\in U(3).
\end{align}\label{pureQLUconditions}
\end{subequations}
For the derivation of these relations, see e.g. \cite{Pati:1983zp,Smirnov:1995jq,Smirnov:2018ske,Perez:2013osa}.

The interactions of $U_1$ conserve baryon and lepton numbers but always introduce lepton flavour violation (LFV) and lepton flavour universality violation (LFUV) -- see \cref{app:LFUV}. 
Hence, the gauge leptoquark is not restricted by proton stability nor by searches for neutrinoless double-beta decay, while extraordinarily high mass limits stem from flavour phenomenology:
assuming $\UL = \UR = \unitmatrix$, the experimental bound  $\BR(K_L^0 \to e^\pm \mu^\mp) < 4.7 \times 10^{-12}$ \cite{Ambrose:1998us} implies $m_{U_1} \gtrsim 2000\,\mathrm{TeV}$. 
However, the gauge leptoquark has different phenomenology with different forms of $\ULR$.

\subsection{Literature overview}

Studies of the $U_1$ leptoquark have gained popularity in recent years as it has been identified as an excellent candidate to account for the neutral-current as well as charged-current $B$-meson anomalies (e.g.~\cite{Kumar:2018kmr,Aebischer:2019mlg}).
The benchmark setup for accommodation of the $B$~ano\-malies as identified in Ref.~\cite{Aebischer:2019mlg} can be written as
\begin{align}
\frac{g_4 \UL}{m_{U_1}} &=\frac{1}{2\,\mathrm{TeV}}\begflmatrix 
    0&0&0\\
    0&-0.05\xi&0.6\\
    0&0.05/\xi&0.7
    \enflmatrix
, &
\frac{g_4 \UR}{m_{U_1}} &=\begflmatrix 0&0&0\\0&0&0\\0&0&0\enflmatrix
,
\label{B-anomalies-solution}
\end{align}
where $\xi$ is a positive $O(1)$ number. 
Clearly, such flavour and chirality pattern is incompatible with the conditions in Eqs.~\eqref{pureQLUconditions}.
For this reason, most of the current studies employ \emph{chiral} vector-LQ models, based on more complicated gauge groups or on complete abandonment of the gauge nature of the $U_1$ field.


Despite its inability to account for the discrepancies in the $B$-meson decays, the gauge leptoquark in the  QLU framework is worth a detailed and dedicated study as it is a common feature of many specific models.
Several top-down studies 
have already been published in the last decades.

In 1994, Valencia and Willenbrock~\cite{Valencia:1994cj} considered the cases where $\UL=\UR$ are permutation matrices, i.e. where each lepton is coupled to a single quark, and studied various two-body meson and tau decays. 
They found that apart from $\KL\to e\mu$, the gauge LQ mass was for some mixing patterns limited from below to 250~TeV by 
$\Ratio_{e/\mu}(\pi^+\to \lep^+ \nu)$ or 
$\Ratio_{e/\mu}(K^+\to \lep^+ \nu)  $, or
 by $\BR(B^+ \to e^+ \nu)$ to $m_{U_1} > 13 \TeV$.
At around the same time, Kuznetsov and Mikheev  \cite{Kuznetsov:1994tt} considered various (semi)leptonic $K$ and $\pi$ decays and the $\mu\to e$ conversion on nuclei, and cast inequalities employing $m_{U_1}$ and elements of quark-lepton mixing matrices, virtually taking the full freedom in the quark-lepton mixing into account, but still tacitly assuming $\UL=\UR$. 
Apart from $\BR(\KL\to e\mu)$ and $\Ratio_{e/\mu}(K^+\to \lep^+ \nu)$, important bounds  have been found to stem also from BR's of $\KL\to \lep^+\lep^-$, $K\to \pi \mu e$ and from coherent $\mu\to e$ conversion on titanium nuclei.
Needless to say, both analyses \cite{Valencia:1994cj} and \cite{Kuznetsov:1994tt}  are outdated nowadays due to  new experimental data.

Concerning more recent works, Ref. \cite{Smirnov:2007hv} considered $\KL\to e\mu$ and $B^0\to e \tau$ for general forms of $\ULR$ but did not confront the obtained limits with other measurements.
In Ref.~\cite{Kuznetsov:2012ai}, which is the 2012 update of \cite{Kuznetsov:1994tt}, also the  $B$ factory results on $B$ and $\tau$ decays have been included and the general case $\UL\neq \UR$ has been considered. 
A specific form of $\UL$ and $\UR$ has been found for which the stated LQ mass limit was as low as 38~TeV.
However, as pointed out in Ref.~\cite{Smirnov:2018ske}, this finding is invalid because the authors forgot to include the predictions for the $\mu^-e^+$ final state
when studying
$\BR(B^0\to \mu^\pm e^\mp)$ and $\BR(B_s \to \mu^\pm e^\mp)$.

Finally, Smirnov~\cite{Smirnov:2018ske} considered all kinematically allowed leptonic decays $P^0 \to \lep^+ \lep^{\prime -}$ for $P^0 = \KL, B^0, B_s$ and took fully into account the freedom in the fermion mixing by performing a scan. 
The global lower limit stemming from these processes was found to be
\begin{align}
m_{U_1} > 86 \TeV
\label{LQmassLimitSmirnov}
\end{align}
and the corresponding forms of $\UL$ and $\UR$ were given. 
We  have verified the computations by completely recalculating  Ref.~\cite{Smirnov:2018ske}.

\subsection{Outline of our work}
\label{outline}
The main goal of this work is to identify all observables which currently determine the gauge LQ mass limit for some form of $\ULR=(\UL, \UR)$. 
These observables are excellent candidates for future New Physics (NP) signals since even a
small improvement in the precision of their measurement shall explore a yet unexcluded part of the parameter space of the model. 
Hence, we call them possible \emph{first future signals} of the gauge LQ. 

Clearly, this is a more ambitious aim than just finding the global LQ mass limit which is the main result of  Ref.~\cite{Smirnov:2018ske}.


In the analysis, we focus especially on the following:
\begin{itemize}

\item We attempt to take into account \emph{all relevant} observables in which the signal of the gauge LQ in the foreseeable future might be potentially found. 
To this end, we employ the Python package \flavio \cite{flavioDocs} which is capable of calculating predictions for hundreds of observables. 

\item More recent measurements are included. 

\item No ad-hoc assumptions are made on the form of $\ULR$. 
Keeping in mind that there is no physically meaningful measure on the parametric space, the setups which might be labeled as \emph{fine-tuned scenarios} or \emph{small parts of the parameter space} are not dismissed.
\end{itemize}

\Cref{sec:model} describes the model in more detail.
In \cref{sec:method},  the technicalities of   the calculations are presented. In \cref{sec:results}, we present the results and discuss the potential of various relevant forthcoming experiments. 
Then in \cref{sec:extended}, we analyze in a similar manner the $\SUC$ models extended by several generations of left- and/or right-handed leptons.  
We briefly conclude afterwards.
In the three appendices we provide  some additional details concerning the lepton flavour group in LQ models, the physics of the $Z'$ boson, and the optimization of the scanning procedure, respectively.

\section{Model details}
\label{sec:model}

The $\SUC$ gauge symmetry can be realized in a minimal way within the \begin{align}\GQL = \SUC \times \SUL \times U(1)_R\end{align}  gauge group \cite{Smirnov:1995jq,Perez:2013osa}. 
This symmetry might be an intermediate stage of a left-right theory based on 
the Pati-Salam group $\GPS = \SUC \times \SUL \times \SUR$ \cite{Pati:1974yy,Pati:1983zp}. 
The spontaneous symmetry breaking (SSB) of $\GQL$ proceeds in two steps as
\begin{align}
    \label{SSB}
    \GQL \to \GSM \to \SUc \times U(1)_Q
    \,.
\end{align}
The generators $T^1_C, \ldots, T^8_C$ of the unbroken part of the $\SUC$ symmetry form $\SUc$, while the weak hypercharge is given by $Y = \sqrt{2/3}\, T^{15}_C + R$. During the first step of symmetry breaking, massive gauge leptoquark $U_1$ and massive $Z'$ arise; the $W$ and $Z$ bosons acquire mass during the second step in a SM-like manner. For further details we refer to \cite{Smirnov:1995jq,Perez:2013osa} or to \cref{app:Zprime}.

The fermion sector consists of 3 generations of the fields in \cref{fermionsIn4plets}. Independent quark and charged-lepton masses can be achieved by using both 1- and 15-dimensional scalar representation of $\SUC$ \cite{Pati:1983zp,Smirnov:1995jq,Perez:2013osa}. 
Concerning the neutrinos, in principle one can assume that they are of Dirac nature \cite{Smirnov:1995jq,Popov:2005mh,Popov:2005wz,Povarov:2011zz,Frolov:2016gvu}. In such a case, the tiny neutrino masses are obtained as a difference of two parameters of the order of the top-quark mass. 
To avoid such a  fine-tuning, one might call for some form of a seesaw mechanism.
The traditional type-I seesaw (studied recently in Ref.~\cite{Murgui:2021bdy} in this context) would require that the $\SUC$ breaking scale is so high that the gauge LQ would have no measurable low-energy phenomenology; hence, this case is not of interest for us. 
Nevertheless,  unification of quarks and leptons is possible even at a low scale when employing the inverse seesaw \cite{Perez:2013osa,Murgui:2021bdy,Murgui:2021bdy} instead. 
This model has been recently studied in  context of the $B$-anomalies  addressed by the \emph{scalar} leptoquarks by Faber et al. \cite{Faber:2018qon,Faber:2018afz}  and also by  Fileviez P. et al. \cite{FileviezPerez:2021lkq,FileviezPerez:2022rbk} with mutually conflicting conclusions.

In accordance with the inverse-seesaw model, we assume heavy $\nu_\Rhand$ in this study. Nevertheless, as we shortly discuss in \cref{sec:results}, the results would be essentially identical also in the  Dirac-neutrino case.

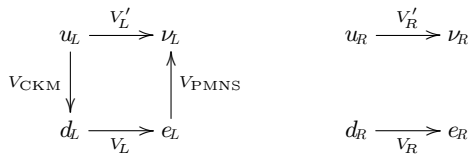
\begin{figure}
    \centering
    \begin{align*}
\xymatrix{
   \uL \ar[r]^\ULp \ar[d]_\CKM 
    	& \nuL  \\
    \dL \ar[r]_{\UL}       & \eL\ar[u]_{\PMNS} }
&&&
\xymatrix{
    \uR \ar[r]^\URp  & \nuR  \\
    \dR \ar[r]_{\UR}       & \eR }  
\end{align*}
    \caption{Scheme of fermion mixing in the quark-lepton symmetry models based on the $\GQL$ gauge group. Here $\ULp = \CKM \,\UL\, \PMNS$.}
\label{fig:fermion_mixings}
\end{figure}

The interplay among the flavour matrices $\UL, \UR, \URp$ introduced in \cref{Lag-U1} and the weak interaction matrices $\PMNS$ and $\CKM$  
is illustrated in \cref{fig:fermion_mixings}.
Adopting the standard (single-phase) parametrization of $\CKM$ and $\PMNS$, no complex phases can be removed from $\UL$ or $\UR$. 
By expanding the $\SUL$ structure in \cref{Lag-U1}, the interactions of the SM fermions with the gauge LQ can be rewritten as
\begin{align}
\Lag_{U_1} &= \frac{g_4}{\sqrt 2}\left(
	\dbar{\hat d} \gamma^\mu
	\left[\PL  \UL	+\PR \UR	\right]
	\hat e
+\dbar{\uLmass} \gamma^\mu
	\CKM \UL  \tilde{\nu}_\Lhand
\right) {U_1}_\mu \nonumber \\ &+ \hc,
\label{Lag-U1-altern}
\end{align}
where $\tilde{\nu}$ is a neutrino flavour triplet in the weak interaction basis and $\PLR = (1\mp \gamma_5)/2$ are chirality projectors.
As the particular form of the $\URp$ matrix is  inconsequential for all the considered low-energy processes, the relevant dimensions of the parameter space are given solely by $2\times9$ angles or phases of $\ULR$ and by $m_{U_1}$.

We do not take into account any other BSM field in the model. 
Especially, the scalar sector is neglected. 
Note that the free parameters of the full renormalizable model \cite{Perez:2013osa} (or  \cite{Smirnov:1995jq}) indeed allow for the regime in which the gauge LQ signals dominate over those of the scalars.
Notice also that the interactions of $Z'$ are flavour-diagonal and hence, its effects in flavour physics are suppressed.
If there is no intermediate stage in the $\GQL \to \GSM$ symmetry breaking,  $m_{Z'}$ is of the same order as $ m_{U_1}$; in such a case, $Z'$ can be also safely neglected. For more details, see \cref{app:Zprime}.

\section{Methods}
\label{sec:method}
In what follows, by \emph{parameter space} we mean the $18$-dimensional set of forms of $\ULR$. 
A \emph{parameter point} is an element of this set
, 
parametrized by angles and phases  $\lambda_{L,ij}$ and $\lambda_{R,ij}$ as described in \cref{app:composite parametrization}.


We have employed two different approaches to investigate a chosen parameter point.
%
The \emph{simplified approach}, adopted from Ref.~\cite{Smirnov:2018ske}  and detailed in \cref{sec:SmirnovsCalculation}, served as a primary 
stage providing basic but yet coherent insight into the parameter space. Within the concept, the identification of interesting parts of the parameter space was 
quite straightforward since it makes use of simple analytical formulae for observable predictions.
The \emph{more robust approach}, described in \cref{sec:robustApproach}, is more comprehensive, but also much less intuitive since it is based on numerical packages which we have used mostly as a black-box tool.

The former approach served also as an important crosscheck which enabled us to find and correct an error in the \flavio package.\footnote{There was a bug in the expression for the $\KLS\to e^\pm \mu^\mp$ amplitudes. 
}
Hence, even tough the  presented results  are based solely on the latter, we still find it worthy to present also the first approach below.

In \cref{sec:scanning}, we describe how the analyzed parameter points have been chosen.

\subsection{Simplified approach}
\label{sec:SmirnovsCalculation}
This approach directly follows Ref.~\cite{Smirnov:2018ske}. 
There are several aspects about this procedure  worth mentioning:
\begin{enumerate}
\item The effects of the $U_1$ leptoquark are taken into account at the tree level.
\item Four-loop QCD running of the induced effective operators is taken into account \cite{privateSmirnov}. For simplicity, the effective operators are defined at the $100~\TeV$ scale, regardless of the considered LQ mass.
\item SM contributions to the considered processes are completely neglected in the calculation. To highlight this approximation, the corresponding predictions for branching ratios are labelled by $\BRV$. 
The measured BR’s of the decays which have been already observed (i.e., $\KL \rightarrow  ee$, $\KL \rightarrow  \mu\mu$, $B^0_s \rightarrow \mu\mu$) are taken as limits on $\BRV$. Such a rough approximation is meaningful due to large relative theoretical uncertainties for the SM amplitudes.
\item Ref. \cite{Smirnov:2018ske} has taken into account the branching ratios of $P \rightarrow l^{\pm} l^{\prime\mp}$ decays where $P = K^0_L$ , $B^0$, $B^0_s$ and $ll^\prime$ corresponds to various kinematically allowed combinations of leptons and antileptons. In our work, also the leptonic decays of $K^0_S$ are considered. The limits on $\Bds \rightarrow  e^\pm\mu^\mp$ are updated \cite{Aaij:2019okb}.
\item No processes with neutrinos are analyzed; the study holds for both situations with light or heavy right-handed neutrinos.
\item The masses of electrons and muons in the final state are neglected, as well as the indirect CP violation in the neutral kaon mass eigenstates.
\item \label{smirnovApproach:firstObs}For given $\ULR$, the LQ mass limit is determined as the maximum of individual limits obtained from the considered observables. The decay responsible for the strongest limit is considered to be the candidate for the \emph{future first signal} of the LQ for the investigated form of $\ULR$.
\end{enumerate}

\begin{figure}
    \centering
    \includegraphics[scale=0.85]{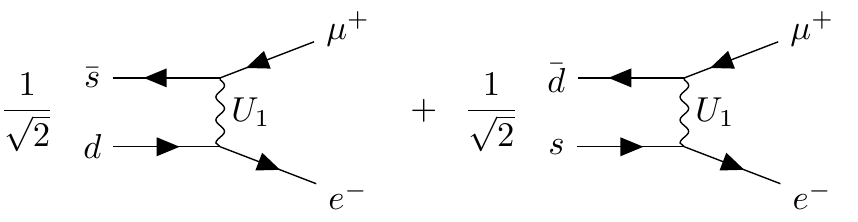}
    \caption{A tree level LFV decay of $\KL$.}
    \label{fig:K0L}
\end{figure}

The branching ratio for a decay with light leptons only is calculated by the following formula:
\begin{equation}\label{brv}
\BRV (P \rightarrow l^+l^{\prime -} ) =\frac{m_P \pi \alpha^2_s f^2_P \overline{m}^2_P (R^V_P )^2}{2m^4_{U_1} \Gamma^{tot}_P} \beta^2_{P,ll^\prime } ,
\end{equation}
where the formfactors are $f_K = 155.72 \MeV$, $f_{B^0} = 190.9 \MeV$, $f_{B^0_s} = 227.2 \MeV$ and $\overline{m}_P = m_P^2/(m_{\bar q}+m_q)$ with $\overline{q}$ and $q$ standing for the index of the valence antiquark and quark of $P$, respectively. 
The gluonic corrections to the pseudoscalar quark currents amount to $R^V_K = 3.47$ and $R^V_B = 2.1$ \cite{privateSmirnov}. 
The lepton-flavour-dependent factor is a sum over two different helicity combinations
\begin{equation}
\beta^2_{P,ll^\prime }=\frac{|a_{LR}(P,l,l^\prime)|^2+|a_{RL}(P,l,l^\prime)|^2}{2},
\end{equation}
where for weak eigenstates
\begin{subequations}
\begin{align}
a_{LR}(P,l,l^\prime) &= \Ul{\overline{q}l} \, \Ur{ql^\prime}^{*} ,\\
a_{RL}(P,l,l^\prime) &= \Ur{\overline{q}l} \,\Ul{ql^\prime}^{*},
\end{align}\label{eigen}
\end{subequations}
while for the CP eigenstates,
\begin{equation}\label{KS,KL}
a(K^0_{L,S},l,l^\prime)=\frac{a(K^0,l,l^\prime)\pm a(\overline{K}^0,l,l^\prime)}{\sqrt{2}}.
\end{equation}
Here $+$ and $-$ relate to $K^0_L$  and $K^0_S$, respectively, and $a $ stands for either $a_{LR}$ or $a_{RL}$. See \cref{fig:K0L} for an illustration.

For processes with a single $\tau$ lepton in the final state, the expression for $\BRV$ in \cref{brv} must be multiplied by a phase space factor \mbox{$(1-m^2_\tau/ m^2_P )^2$}. 
Along with that, the replacement 
$\UlR{r\tau}\rightarrow \left[ \UlR{r\tau}-\UrL{r\tau}  m_\tau /(2\overline{m}_P R^{V}_P )  \right]$
for $r = q$, $\bar{q}$ is applied in \cref{eigen}. 
For $\tau^+\tau^-$ in the final state see Ref.~\cite{Smirnov:2018ske}. 


\subsection{More robust approach} \label{sec:robustApproach}
In parallel with the previous approach, we have also performed a similar analysis using the family of general-purpose open-source tools \wilson \cite{Aebischer:2018bkb,wilsonDocs}, \flavio \cite{Straub:2018kue,flavioDocs}, and \smelli \cite{Aebischer:2018iyb,smelliDocs}. 
We present the features of this approach as a list which can be compared with that in the previous section.
\begin{enumerate}
\item The  LQ  interactions are matched onto the Standard Model effective field theory (SMEFT) at the tree level (similarly to the previous approach), yielding  non-zero Wilson coefficients
\begin{subequations}\label{Wilsons-GLQ}
\begin{align}
C_{ed\,\_\,\bar\lep\lep\bar qq }
&=-1\frac{g_4^2}{2m_{U_1}^2}
\Ur{q\bar\lep}^* \,\Ur{\bar q\lep} 
\,,
\\
C_{\ledq\,\_\,\bar\lep\lep\bar qq }
&=+2 \frac{g_4^2}{2m_{U_1}^2} 
\Ur{q\bar\lep}^* \, \Ul{\bar q\lep} \,,
\\
C^{(1)}_{\LeptonDoublet\QDoubletIndex\,\_\,\bar\lep\lep\bar qq }=
C^{(3)}_{\LeptonDoublet\QDoubletIndex\,\_\,\bar\lep\lep\bar qq}
&=-\frac{1}{2}\frac{g_4^2}{2m_{U_1}^2}
\Ul{q\bar\lep}^* \, \Ul{\bar q\lep} \,,
\end{align}
\end{subequations}
which multiply the following effective operators (with flavour indices suppressed):
\begin{subequations} \label{GLQ-SMEFT-op}
\begin{align}
\Op_{ed}&=(\dbar{\eR} \gamma_\mu \eR)(\dbar{\dR}\gamma^\mu \dR)\,,
\\
\Op_{\ledq}&=(\dbar{\LL} \eR)(\dbar{\dR} \QL)\,,
\\
\begin{split}\Op_{\LeptonDoublet\QDoubletIndex}^{(1)} + \Op_{\LeptonDoublet\QDoubletIndex}^{(3)}
&=(\dbar{\LL} \gamma_\mu \LL)(\dbar{\QL}\gamma^\mu \QL)\\
&\;+ (\dbar{\LL} \gamma_\mu \sigma^I \LL)(\dbar{\QL}\gamma^\mu \sigma^I \QL)\,.\end{split}
\end{align}
\end{subequations}

We have implemented a python function taking $\ULR$ and $m_{U_1}$ as input arguments and returning a dictionary of SMEFT Wilson coefficients in the format compatible with the \wcxf standard \cite{Aebischer:2017ugx,wcxfDocs}, which is used by the packages mentioned above.

\item The renormalization group (RG) running of the SMEFT effective operators from the scale $\mu = m_{U_1}$ to the electroweak scale, the tree-level matching onto the Weak effective theory (WET) and further evolution to the meson-mass energy scales is handled automatically by the \wilson package. 
The full numerical solution to the one-loop SMEFT RG equations (the \texttt{'integrate'} option) is performed since we have exemplified that the \texttt{'leadinglog'} approximation leads to $O(1)$ relative differences in certain predictions. 
 Analytical solution to the one-loop QCD and QED running equations is applied under the electroweak scale in \wilson. 
For more details see \cite{Aebischer:2018bkb} and references therein.

\item The SM contributions to the amplitudes of the calculated processes are automatically taken into account by \flavio. 
As a result of this (and of the RG running), the predictions  do not  scale  uniformly as $m_{U_1}^{-4}$, which was a simplifying feature of the previous approach [see~\cref{brv}]. 

\item  
The global likelihood tool \smelli is employed. This package uses \flavio for predictions and confronts them with the measurements, including correlations. 
By default, version \texttt{2.2.0} of \smelli takes into account hundreds of observables, most of which are, however, irrelevant for our scenarios.
On the other hand, the very interesting processes  $\BR(\Bds\to e^+e^-)$ as well as $\mu\to e$ conversion on nuclei were not included. 
To this end, we have modified the \smelli package to calculate also these observables.

The complete list of considered observables can be found in \cite{Hudec:2021zpr} or inferred from \cite{smelliGithub}.

\item No light right-handed neutrinos are assumed.

\item Light lepton masses are taken into account in \flavio for all observables, but indirect CP violation in neutral kaons remains neglected in the $\KLS\to \lep \lep'$ decays.

\item \label{more robust how first signal is obtained} For $\UL$ and $\UR$ fixed, we find  $m_{U_1}$ for which the global log-likelihood calculated by \smelli 
worsens by 4 units with respect to the SM. That value defines the lower LQ mass limit for this particular case.
Then, the corresponding candidate for the \emph{future first signal} of NP is the observable for which the individual pull between theory and experiment worsened the most compared to the SM case; the pulls have been obtained via the \texttt{obstable} method provided by the \smelli package \cite{Aebischer:2018iyb}.

We have also tried different (more complicated) criteria, supposed to underpin  scenarios in which the likelihood actually improves, but we ended up with qualitatively identical results.

\end{enumerate}

\subsection{Analyzing the parameter space}
\label{sec:scanning}

\begin{figure}
\includegraphics[scale=0.8]{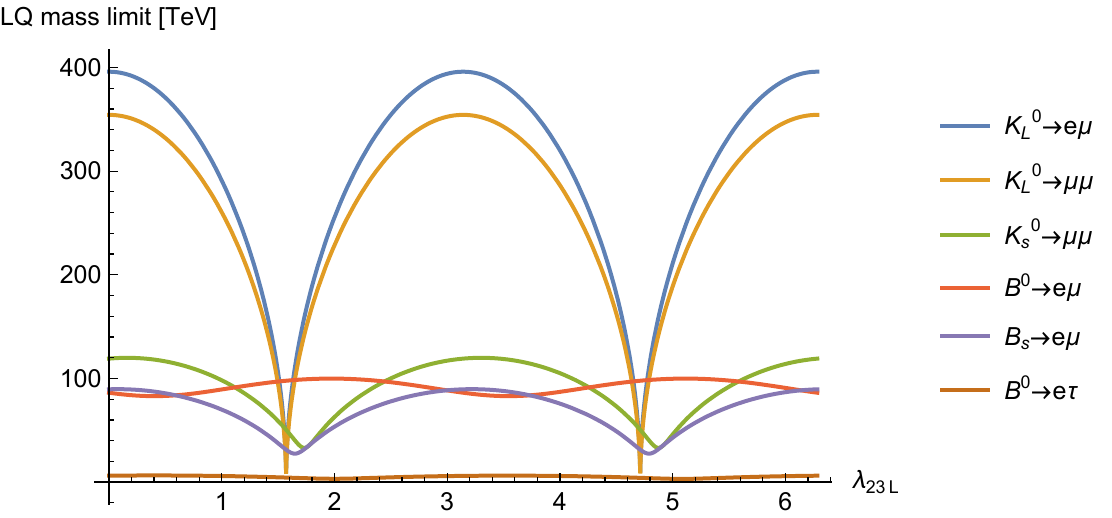}
\caption{Illustration of lower limits on the gauge LQ mass stemming from several observables, calculated along a one-dimensional cut of the parameter space described in \cref{app:composite parametrization}. On this slice of the parameter space, the bounds are given by $\BR(\KL\to e^\pm\mu^\mp)$ and $\BR(B^0\to e^\pm \mu^\mp)$.
The mass limits are obtained  using the approach described in \cref{sec:SmirnovsCalculation}.}
\label{fig:picr}
\end{figure}

The final analysis has been performed within the \emph{more robust approach} where analyzing a single parameter point typically takes over a minute on a usual computer.
Apparently, an 18-dimensional 
parameter space cannot be rigorously explored just with a blind numerical scan. 
%
To this end, we have addressed the issue in two mutually complementary 
ways:

\begin{itemize}

\item
A series of random numerical scans has been performed, using a naïve measure $\Pi_{ij}\mathrm{d}\lambda_{L,ij} \mathrm{d}\lambda_{R,ij}$, where $i,j$  run only over the unfixed $\lambda$'s. The gradual fixing of $\lambda$'s proceeded along the following lines (the details can be found in \cref{app:composite parametrization}):


In the first stage, $10^3$ parameter points have been obtained 
with none of the $\lambda$'s fixed.
In   majority of cases, the limiting processes were $\BR(\KL\to e\mu)$, $\BR(\KL\to ee)$
 and $\CRmueAu$, i.e., the coherent conversion rate of $\mu\to e$ on nuclei.

 
 Following Ref.~\cite{Smirnov:2018ske} in the second stage, about $2\times 10^3$ parameter points have been obtained 
 on the parameter subspace defined by $\BRV(\KL\to ll^{\prime})=0$, achieved by \cref{nes condition,general solution}. This is motivated by exploring the "steep valleys" on \cref{fig:picr}. 
Now $\CRmueAu$ dominated almost all cases. 
 
 In the third stage, more than $10^4$ parameter points have been obtained by random scanning on the parameter subspace restricted both by $\BRV(\KL\to ll^{\prime})=0$ and $\CRmueAu\approx0$, i.e., by Eqs. \eqref{nes condition}, \eqref{general solution}, and \eqref{avoidmue}.


\item
%
We have compiled a list of relevant observables discussed in the recent review in Ref.~\cite{Dorsner:2016wpm} and investigated if they might become the \emph{future first signal}. 
For each of these observables, we have found either a  parameter point for which this observable is the first future signal indeed
, or an argument that such a point should not exist.
A thorough effort has been made to include various special parts of the parameter space in the considerations.

\end{itemize}


Combining those two methods enables us to claim with a higher level of confidence that the catalogue in \cref{tab:LimitingBRs} is \emph{complete}.

\section{Results}
\label{sec:results}

\begin{samepage}
    
\begin{figure*}[p]
    \centering
    \includegraphics{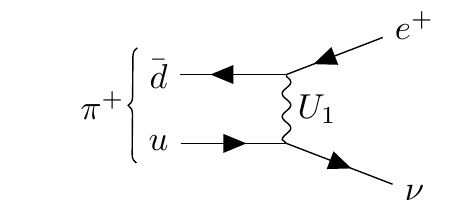}
    \quad
    \includegraphics{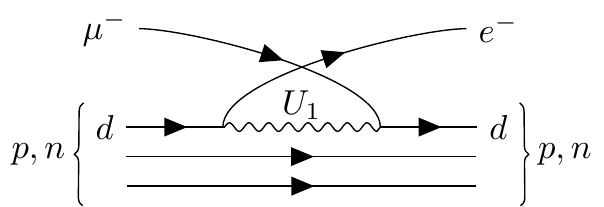}
    \quad
    \includegraphics{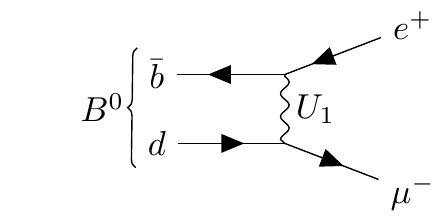}
    \caption{Examples of Feynman graphs underpinning the possible first signals of the $U_1$ gauge leptoquark.}
    \label{fig:my_label}
\end{figure*}

\begin{table*}[p]
\centering
\rowcolors{1}{green!80!blue!6}{white}
\begin{tabular}{>{$}l<{$}  >{{\hspace{10px}}$}r<{$}  l<{\hspace{10px}} >{$}r<{$}  l }
\hline
\rowcolor{gray!40!green!30}
\text{Observable}
	& \multicolumn{1}{r}{Experiment} &
	& \multicolumn{1}{r}{SM prediction}  &\\ \hline
\BR(\KL\to e^\pm \mu^\mp )
	& < 4.7 \times 10^{-12} & \cite{Ambrose:1998us} 
	& 0 & \\
\BR(\KL \to e^+ e^- )
	&  8.7^{+5.7}_{-4.1} \times 10^{-12} 
		& \cite{Ambrose:1998cc}
	& (9.0\pm 0.5)\times 10^{-12}
		& \cite{Valencia:1997xe,GomezDumm:1998gw}\\
\BR( \KL \to \mu^+\mu^- )
	& (6.84 \pm 0.11) \times 10^{-9} 
		& \cite{Zyla:2020zbs} 
	& (7.4 \pm 1.3) \times 10^{-9}\\
\BR(\KS \to \mu^+\mu^- )
	& < 2.1 \times 10^{-10} 
		& \cite{Aaij:2020sbt}
	& (5.2 \pm 1.5) \times 10^{-12}
		& \cite{DAmbrosio:2017klp} \\
\BR(B^0 \to  e^\pm \mu^\mp )
	& < 1.0\times 10^{-9} 
		& \cite{Aaij:2017cza} 
	& 0 &\\
\BR(B_s \to  e^\pm \mu^\mp )
	& < 5.4\times 10^{-9}
		& \cite{Aaij:2017cza}
	& 0 & \\
\BR(B^0 \to  \mu^+\mu^- )
	& 1.1^{+1.4}_{-1.3} \times 10^{-10} 
		& \cite{Zyla:2020zbs}
	& (1.1 \pm 0.1) \times 10^{-10}	& \\
\BR(B_s \to  \mu^+\mu^-) 
	& (3.0 \pm 0.4) \times 10^{-9} 
		& \cite{Zyla:2020zbs}
	& (3.7 \pm 0.2) \times 10^{-9}& \\ 
\Ratio_{e/\mu}(\pi^+ \to  \lep^+ \nu)
	&  1.2327(23) \times 10^{-4}
		& \cite{Zyla:2020zbs}
	& 
	  1.2352(1)\times 10^{-4} 
		& \cite{Cirigliano:2007ga}\\
\Ratio_{e/\mu}(K^+ \to \lep^+ \nu)
	& 
		2.488(9)\times 10^{-5} 
		& \cite{Zyla:2020zbs}
	& %
	 2.476(2)\!\times\! 10^{-5} &
\\
\CRmueAu
 & < 7 \times 10^{-13} & \cite{Bertl:2006up}
 & 0  & 
 \\ \hline
\end{tabular}
\caption{Complete list of observables which currently constrain the gauge LQ mass for some  form of the $\ULR$ matrices. 
The experimental limits are given   at 90\% C.L.
The SM predictions have been calculated in \flavio unless cited.}
\label{tab:LimitingBRs}
\end{table*}

\begin{table*}[p]
\centering
\rowcolors{1}{yellow!80!blue!13}{white}
\begin{tabular}{
>{$}l<{$} 
>{$}r<{$}  l
>{$}r<{$} 
>{$}r<{$}  l}
\hline
\rowcolor{gray!20!yellow!30}
\text{Observable} 
	& \multicolumn{1}{r}{Experimental limit} 
	& &\multicolumn{1}{r}{QLU model prediction}
	& \multicolumn{1}{r}{\phantom{xx}SM prediction}  & \\ \hline
\BR(\KS \to e^+ e^-)
 & < 9 \times 10^{-9} & \cite{Ambrosino:2008zi}
 & \leq 2 \times 10^{-9}
 & 2 \times 10^{-14} & \cite{Ambrosino:2008zi}\\
\BR(\KS \to e^\pm \mu^\mp)
 & \text{N/A} & \cite{Zyla:2020zbs}
 & \leq 3 \times 10^{-10}
 & 0 &\\
\BR(B^0 \to e^+ e^-)
 & < 2.5\times 10^{-9} & \cite{Aaij:2020nol}
 & \leq 1.1 \times 10^{-10}
 & 3 \times 10^{-15} & \cite{Bobeth:2013uxa}\\
\BR(B_s \to e^+ e^-) 
 & < 9.4\times 10^{-9} & \cite{Aaij:2020nol}
 & \leq 3 \times 10^{-9}
 & 9 \times 10^{-14} & \cite{Bobeth:2013uxa}\\
\BR(B^0 \to e^\pm \tau^\mp)
 & < 2.8\times 10^{-5} & \cite{Aubert:2008cu}
 & \leq 6 \times 10^{-9}
 & 0 &\\
\BR(B_s \to e^\pm \tau^\mp)
 & \text{N/A} & \cite{Zyla:2020zbs}
 & \leq 2.5 \times 10^{-9} 
 & 0 &\\
\BR(B^0 \to \mu^\pm \tau^\mp)
 & < 1.2\times 10^{-5}  & \cite{Aaij:2019okb}
 & \leq 5 \times 10^{-9}
 & 0 &\\
\BR(B_s \to \mu^\pm \tau^\mp)
 & < 3.4\times 10^{-5} & \cite{Aaij:2019okb}
 & \leq 2.3 \times 10^{-9}
 & 0 &\\
\BR(B^0 \to \tau^+ \tau^-)
 & < 1.6 \times 10^{-3} & \cite{Aaij:2017xqt}
 & 2 \times 10^{-8}
 & 2 \times 10^{-8} & \cite{Bobeth:2013uxa}\\
\BR(B_s \to \tau^+ \tau^-)
 & < 5.2\times 10^{-3} & \cite{Aaij:2017xqt}
 & 8 \times 10^{-7} 
 & 8 \times 10^{-7} & \cite{Bobeth:2013uxa}
\\ \hline
\end{tabular}
\caption{Examples of processes which are not listed in \cref{tab:LimitingBRs}. The third column shows predictions 
 obtained during  the numerical scanning 
 following from the forms of  $\UL, \UR$, and $m_{U_1}$
 which are fully compatible with all the current experimental limits. We also list the SM predictions for comparison.}
\label{tab:VLQ-otherProcesses}
\end{table*}

\end{samepage}

\begin{table*}[p]
\centering
\begin{tabular}{LCCC}
\hline
\rowcolor{brown!20}
\text{Observable} & \UL & \UR & \text{Limit on }m_{U_1} 
\\[1pt]  \hline \noalign{\vskip 4pt}
\BR(\KL\to e \mu)
& \begflmatrix
 1 & 0 & 0 \\
 0 & 1 & 0 \\
 0 & 0 & 1 
 \enflmatrix 
 & \begflmatrix
 1 & 0 & 0 \\
 0 & 1 & 0 \\
 0 & 0 & 1 
\enflmatrix
& 2074\TeV
\\[5mm]  \arrayrulecolor{brown!20}\hline\noalign{\vskip 4pt}
\BR(\KL\to e e)
 & \begflmatrix
 \frac{1}{\sqrt{2}} & 0 & \frac{1}{\sqrt{2}} \\
 \frac{1}{\sqrt{2}} & 0 & -\frac{1}{\sqrt{2}} \\
 0 & 1 & 0 
\enflmatrix
 & \begflmatrix
 \frac{1}{\sqrt{2}} & 0 & \frac{1}{\sqrt{2}} \\
 \frac{1}{\sqrt{2}} & 0 & -\frac{1}{\sqrt{2}} \\
 0 & 1 & 0 
 \enflmatrix 
 & 1335\TeV
\\[6mm] \hline\noalign{\vskip 2pt}
 \BR(\KL\to \mu \mu)
 & \begflmatrix
   0 & \frac{1}{\sqrt{2}} & \frac{1}{\sqrt{2}} \\
   0 & \frac{1}{\sqrt{2}} & -\frac{1}{\sqrt{2}} \\
   1 & 0 & 0 
   \enflmatrix 
 & \begflmatrix
   0 & \frac{1}{\sqrt{2}} & \frac{1}{\sqrt{2}} \\
   0 & \frac{1}{\sqrt{2}} & -\frac{1}{\sqrt{2}} \\
   1 & 0 & 0 
   \enflmatrix
 & 319\TeV 
\\[5mm] \hline\noalign{\vskip 2pt}
 \BR(\KS\to \mu \mu)
 & \begflmatrix
 0 & \frac{1}{\sqrt{2}} & \frac{1}{\sqrt{2}} \\
 0 & \frac{1}{\sqrt{2}} & -\frac{1}{\sqrt{2}} \\
 1 & 0 & 0 
 \enflmatrix 
 &
\begflmatrix
 0 & \frac{1}{\sqrt{2}} & \frac{1}{\sqrt{2}} \\
 0 & -\frac{1}{\sqrt{2}} & \frac{1}{\sqrt{2}} \\
 1 & 0 & 0 
\enflmatrix
 & 153\TeV
\\[6mm] \hline\noalign{\vskip 2pt}
 \BR(B^0\to \mu\mu)
&\begflmatrix 
 0 & -\frac{1}{\sqrt{21}} & -\sqrt{\frac{20}{21}} \\
 0 & - \sqrt{\frac{20}{21}} & \frac{1}{\sqrt{21}} \\
 -1 & 0 & 0
 \enflmatrix & \begflmatrix 
 -\frac{1}{\sqrt{21}} & 0 & -\sqrt{\frac{20}{21}} \\
 \sqrt{\frac{20}{21}} & 0 & -\frac{1}{\sqrt{21}} \\
 0 & -1 & 0 
  \enflmatrix 
& 102\TeV
\\[6mm]  \hline\noalign{\vskip 2pt}
\BR(B_s\to \mu\mu)
&
\begflmatrix
  0. & 0. & i \\
 -0.26-0.34 i & 0.78\, -0.45 i & 0. \\
 -0.74-0.52 i & -0.29+0.32 i & 0. \\
\enflmatrix
&\begflmatrix
 0. & 0. & 1\\
 0.20\, -0.29 i 
 		& 0.83\, -0.43 i & 0. \\
 -0.14-0.92 i & -0.12+0.34 i & 0. 
\enflmatrix
& 290\TeV
\\[6mm] \hline\noalign{\vskip 2pt}
\BR(B^0 \to e\mu)
 & \begflmatrix
  0 & -1 & 0 \\
  0 & 0 & 1 \\
  -1 & 0 & 0 
  \enflmatrix 
 & \begflmatrix
  0 & -1 & 0 \\
  0 & 0 & 1 \\
  -1 & 0 & 0 
  \enflmatrix 
 & 123\TeV
\\[6mm] \hline\noalign{\vskip 2pt}
\BR( B_s \to e\mu )
 & 
  \begflmatrix
   0 & -0.04-0.06i & -0.09-0.99i \\
   0 & 0.20-0.98i & -0.05+0.06i \\
   1 & 0 & 0 
  \enflmatrix 
 & \begflmatrix
   0 & -0.06+0.04i & -0.23-0.97i \\
   0 & 0.12-0.99i & -0.06-0.04i \\
   1 & 0 & 0 
  \enflmatrix 
 & \begin{array}{c} 90\TeV \\ \text{(global limit)}\end{array}
 \\[7mm] \hline\noalign{\vskip 2pt}
 \Ratio_{e/\mu}(K^+ \to \lep \nu)
 & \begflmatrix
   0 & 0 & 1 \\
   1 & 0 & 0 \\
   0 & 1 & 0
   \enflmatrix 
 & \begflmatrix
   0 & 0 & 1 \\
   1 & 0 & 0 \\
   0 & 1 & 0
   \enflmatrix 
 & 245\TeV
\\[6mm] \hline\noalign{\vskip 2pt}
\Ratio_{e/\mu}(\pi^+ \to \lep \nu)
 & \begflmatrix
  1 & 0 & 0 \\
  0 & 0 & 1 \\
  0 & 1 & 0
  \enflmatrix 
 & \begflmatrix
  1 & 0 & 0 \\
  0 & 0 & 1 \\
  0 & 1 & 0
  \enflmatrix   
 & 270 \TeV
\\[6mm] \hline\noalign{\vskip 2pt}
 \CR( \mu\to e , \mathrm{Au})
& \begflmatrix
   0.38 & 0    & 0.93 \\
   -0.93 i & 0 & 0.38 i \\
 0 &  i & 0 \\
\enflmatrix
&
 \begflmatrix
 0.26 & 0.27 & 0.93 \\
 -0.64 i & -0.67 i & +0.38 i \\
  -0.72 i & +0.69 i & 0
\enflmatrix
& 585 \TeV
\\[6mm] 
 \arrayrulecolor{black}
\hline  
\end{tabular}
\caption{Examples of quark-lepton mixing matrices and the corresponding dominant signals of the gauge leptoquark.}
\label{tab:ULURexamples}
\end{table*}


\begin{samepage}
\Cref{tab:LimitingBRs,tab:ULURexamples} present the catalogue of observables which currently give the most stringent constraint on $m_{U_1}$ for some configuration of $\ULR$. These observables correspond to the \emph{future first signals} as defined above.
To fully appreciate the result, notice that even a very small improvement in precision of any experimental limit listed in \cref{tab:LimitingBRs} will probe a so-far allowed part of the parameter space of the model, and could potentially detect a NP signal --
the only exception is the observed decay $\KL\to\mu\mu$ for which the theoretical uncertainties within the SM dominate.

\end{samepage}

Conversely, under a very idealized assumption that the experimental sensitivity will grow uniformly for all the observables considered, no other observable could become the first observed signal of the gauge LQ. 
More realistically, the measurement  precision of 
any other observable needs to be improved by a larger step in order to put a new constraint on the model parameters or to have a theoretical chance of observing a signal of the gauge LQ.
How large these steps must be is shown for several important examples in \cref{tab:VLQ-otherProcesses}.

\subsection{\texorpdfstring{Global mass limit -- comparison with Ref.~\cite{Smirnov:2018ske}}{Global mass limit -- comparison with Ref.[4]}}
\label{sec:global}

As noted earlier, the 
\emph{simplified approach} of Ref.~\cite{Smirnov:2018ske} described in \cref{sec:SmirnovsCalculation}  
leads to the global lower leptoquark mass limit of $86\TeV$. The corresponding $\ULR$  
is shown in the last line of \cref{tab:ULURexamples}.
However, when taking into account more observables in the \emph{more robust approach}, $m_{U_1}=86\TeV$ for this parameter point
turns out to be in conflict with the bound on $\CRmueAu$ by 3 orders of magnitude.

Nevertheless, we have found a form of $\ULR$ which allows  essentially the same mass (90~TeV, see~\cref{tab:ULURexamples}) even when all the constraints included in \smelli are considered.

\subsection{Possible first signals}
\label{sec:first signals}

Concerning searches for LFV, 
\Cref{tab:LimitingBRs} contains limits on $\KL, \Bds \to e\mu$ and on the $\mu\to e$ coherent conversion on nuclei; further searches for these processes are therefore of great interest. The remaining observables in \cref{tab:LimitingBRs} are all related to the leptonic decays of pseudoscalar mesons which are chirality suppressed in the SM and can be understood as \emph{tests of LFUV in the SM}:

Firstly, significant deviations could arise in the ratios of charged current decays  $R_{e/\mu}(P^+\to\lep \nu)$ with $P = \pi, K$ when the LQ couples mostly to the electrons.
Although the decay widths involved cannot be measured with the precision similar to the rare decays above, the deviations from the SM can be significant due to the interference among the NP and SM amplitudes.
Subdominant contributions arise also from  the other neutrino species as well as from the $\lep_\Lhand \nu_\Rhand$ final state if the right-handed neutrinos are light enough.


Secondly, limits on $m_{U_1}$ stem also from the observed BRs of $\KL \to ee,\mu\mu$ and $\Bds \to \mu\mu$. 
Concerning $\KL \to \mu\mu$, the experimental precision is better than the theoretical error estimates in the SM stemming from long-distance contributions \cite{Isidori:2003ts,Chobanova:2017rkj}.

Finally, a very interesting limit on the $U_1$  mass for some patterns of quark-lepton mixing is set by the recent \LHCb search for $\KS\to\mu\mu$; the anticipated discovery of this decay after the upcoming LHC runs thus provides an exciting opportunity for the Pati-Salam-type leptoquark.

\subsection{Other observables}
\label{sec:second signal}

In \cref{tab:VLQ-otherProcesses},  the  $P^0\to \lep\lep'$ decays that currently do not pose the most stringent bound on $m_{U_1}$ are listed, together with the predictions based on the parameters fully compatible with all the current experimental searches. All generated parameter points  have been included. 

As $\tau$ leptons are generally experimentally hard to handle, all processes involving $\tau$'s belong to this category. 
In fact, $3\sim 4$ orders of magnitude improvements in limits on $\Bds\to \lep \tau$ would be necessary in order to compete with the other constraints, which is far below the prospected sensitivity of \BELLEII  \cite{Kou:2018nap} and hardly achievable even at \LHCb at the high-luminosity phase.
Furthermore, as explained in \cref{app:composite parametrization},  due to the unitarity of $\ULR$, the LQ amplitudes mediating of $\Bds\to \tau\tau$ are severely limited by the probes of $\KL\to l l'$ and, thus, our predictions for the former essentially coincide with the SM. 
Hence, the expected sensitivity of \BELLEII at about $10^{-6}$ for $\BR(B^0\to\tau\tau)$ \cite{Cunliffe:2017cox} shall  not be an interesting probe of the considered model.

On the other hand, the experimental sensitivities to $\KS, B^0$, and $B_s$ decays to $e^+e^-$ require less than 1 order of magnitude improvement in order to probe the currently unexplored parts of the parameter space.  
Note that $\BRV(\Bds \to e^+e^-) = \BRV(\Bds \to \mu^+\mu^-)$ is predicted for any parameter point for which $\BRV(\KL\to\lep\lep')=0$ \cite{Smirnov:2018ske}; currently, the muonic channel is measured more accurately. 
However,
when further searches for NP in the $P^0 \to \mu^+\mu^-$ decays become limited by the SM uncertainties, new searches for
$\Bds, \KS \to ee$ will become essential.

No experimental limits on the decay $\KS \to e \mu$ are available \cite{Zyla:2020zbs}. 
Comparing with the current limits on  $\KS \to e e$ \cite{Ambrosino:2008zi} and $\KS \to \mu\mu$ \cite{Aaij:2020sbt}, we reckon  the required experimental sensitivity around $10^{-10}$  for $\KS \to e \mu$ might be reachable by \collabName{KLOE II}  or \LHCb.

Semileptonic decays like $B\to K \mu\mu$ or loop processes such as $\mu\to e \gamma$ might become the dominant signals of \emph{chiral} leptoquaks but not of the gauge LQ in the considered model  as it inevitably introduces sizable Wilson coefficients $C_{\ledq}$ which are experimentally more constrained (see, e.g., 
Ref.~\cite{Becirevic:2016zri}). 

\section{Extended \texorpdfstring{$\boldsymbol{\SUC}$}{SU(4)C} models}
\label{sec:extended}

This part of our work is devoted to more complicated models featuring the vector leptoquark $U_1$. 
Although they could be considered as aesthetically less appealing, such models have been studied thoroughly in the recent years, mainly due to the attempts to accommodate the $B$-meson anomalies. 
Generally, several tricks to circumvent the theoretical requirement of unitarity of $\UL$ and $\UR$
have been suggested in the literature. 
They can be divided into three categories, according to the paradigm abandoned:
\begin{enumerate} 
\item Adding extra generations of fermions while maintaining the gauge symmetry group $G_{421}$ or $G_{422}$ \cite{Calibbi:2017qbu,Balaji:2018zna,Bernigaud:2021fwn,Balaji:2019kwe}.
\item Assuming more complicated gauge structure. 
Especially, the models based on the 
$G_{4N21} = SU(4)_{C_L} \times SU(N)_{C_R} \times SU(2)_L\times U(1)_R$ gauge symmetry have become popular; here $N=3$ or $4$ and the QCD generators are given by $T^A_C = T^A_{C_L}  + T^A_{C_R}$ for $A=1, \ldots, 8$. 
In the basic setting of \emph{chiral quark-lepton symmetry} \cite{Assad:2017iib,Fornal:2018dqn}, the left-handed fermions are charged by $SU(4)_{C_L}$ while the right-handed ones transform non-trivially under $SU(N)_{C_R}$. Hence, the $U_1^\mu$ field interacting with the left-handed quark-lepton currents is a \emph{chiral leptoquark} -- it has no or suppressed couplings to the right-handed currents, avoiding the scalar-type effective operators $\Op_{\ledq}$ which are responsible for all the most stringent limits in \cref{tab:LimitingBRs}.

In more general cases with $N=3$, some quark and lepton fields are unified within the $SU(4)$ factor while others live in separate irreps of $SU(3)$ \cite{Baker:2019sli}.
 Usually, more than 3 generations of fermions are considered 
\cite{Diaz:2017lit,DiLuzio:2017vat,DiLuzio:2018zxy,Greljo:2018tuh,Cornella:2019hct}.

 For even more exotic gauge groups see, e.g.,~ 
 \cite{Bordone:2018nbg, FernandezNavarro:2022gst}.
%

\item Assuming that the vector LQ is not a gauge field but a composite resonance formed by some more fundamental strongly interacting fields \cite{Barbieri:2015yvd,Barbieri:2016las,Barbieri:2017tuq}.
\end{enumerate}
This work is focusing solely on the first option. 
Since the SM leptons do not entirely stem from the same $\SUC$ representations as the quarks, we shall not use the term \emph{quark-lepton unification} for these theories  but rather call them \emph{extended $\SUC$ models}.

\subsection{Specification of the models}
Like in the previously considered $\SUC\times\SUL\times U(1)_R$ scenarios, see \cref{fermionsIn4plets},  the models contain 3 generations of each of the following  chiral fermion $\SUC$ quadruplets:
\begin{subequations}
\begin{align}
{\FQL}_{(4,2,0)}&=
\begin{pmatrix} \,\QL \\ ^4\LL \end{pmatrix}\,,
\\
{\fuR}_{(4,1,+\sfrac 12)}&=
\begin{pmatrix} \uR\\ {}^4\nuR \end{pmatrix}\,,
\\
{\fdR}_{(4,1,-\sfrac 12)}&=
\begin{pmatrix} \dR\\ ^4\eR \end{pmatrix}\,.
\end{align}\end{subequations}
Notice that we have slightly updated the notation by adding an "isotopic index" to the leptons living inside the quadruplets.
On top of that,  
$\nLext$ 
generations of  $\SUL$-doublet vector-like fermions
\begin{align}
 {}^1\LL_{(1,2,+\sfrac12)} 
+{}^1\LR_{(1,2,+\sfrac12)}
\end{align}
and $\nRext$ generations of weak-singlet vector-like  fermions
\begin{align}
  {}^1\eL_{(1,1,-1)} 
+ {}^1\eR_{(1,1,-1)}
\,,
\end{align}
are assumed. Being $\SUC$ singlets, these new fields are intact to  interactions of the gauge LQ.
After the $\GQL\to\GSM$ symmetry breaking, they can mix with the leptons from the quadruplets.
We assume that the 3 lightest eigenstates correspond to $e,\mu$, and $\tau$, while the $\nLext + \nRext$ remaining ones are too heavy to be observed. 
As the weak hypercharges of the 3 known leptons are quite precisely measured, they must be composed solely from the fields ${}^1\eR, {}^4\eR, {}^1\LL$ and ${}^4\LL$. 
For all practical purposes, it is sufficient to assume the following mixing pattern in the charged-lepton sector:
\begin{align}
\begflmatrix
\eRmass \\ E_\Rhand \\  {}^1\LR^-
\enflmatrix
 = 
\begflmatrix
	{\VeR}
&
	\begin{matrix}
		0_{3\times \nLext}\\
		0_{\nRext\times \nLext}
	\end{matrix}
\\
	\begin{matrix}
		0_{\nLext\times 3}
		&0_{\nLext\times \nRext}
	\end{matrix}
&
\unitmatrix_{\nLext\times\nLext}
\enflmatrix
\begflmatrix
	{^4\eR} \,
\\
	{^1\eR} \,
\\
{}^1\LR^-
\enflmatrix,
\\   
\begflmatrix
\eLmass \\ E_\Lhand \\  {}^1\eL
\enflmatrix
 = 
\begflmatrix
	{\VeL}
&
	\begin{matrix}
		0_{3\times \nRext}\\
		0_{\nLext\times \nRext}
	\end{matrix}
\\
	\begin{matrix}
		0_{\nRext\times 3}
		&0_{\nRext\times \nLext}
	\end{matrix}
&
\unitmatrix_{\nRext\times\nRext}
\enflmatrix
\begflmatrix
	{^4\LL^-}
\\
	{^1\LL^-}
\\
{}^1\eL
\enflmatrix,
\end{align}
where, generally,  $\LeptonDoublet^-$ denotes the electrically charged component of an $\LeptonDoublet$ doublet %
 (notice that $^1\eL \neq {}^1\LL^-$ and $^1\eR \neq {}^1\LR^-$), \mbox{$\hat e = \eLmass + \eRmass$} is the triplet of light leptons while $E_\Rhand$ and $^1\LR^-$ with their chiral counterparts $E_\Lhand$ and $^1 \eL$ form the heavy mass eigenstates. 
The  form of the mixing in the heavy-lepton sector is irrelevant for our considerations. 
The blocks $\VeL$ and $\VeR$ are arbitrary unitary matrices of dimension $3+\nLext$ and $3+\nRext$, respectively.
Including the "non-standard" fields $^1\LR$ and $^1\eL$ into the model ensures the ABJ anomaly cancellation and enables one to write down arbitrarily large Dirac mass terms for the vector-like pairs.

The $Q=0$ components of $^4\LL$ and $^1\LL$ naturally follow their charged $\SUL$ partners during the mixing at the first stage of SSB: those belonging to $E_\Lhand$ become equally heavy while the companions of $\eLmass$ become the light neutrinos, eventually gaining mass after the electroweak symmetry breaking.

There are no extra quarks in the models and the transformation from gauge to mass eigenstates is given by $3\times3$ unitary matrices: 

\begin{subequations}
\begin{align}
\massbasis{u}_{\Lhand} &= V^u_{L} \, u_{\Lhand}
\,, &
\massbasis{u}_{\Rhand} &= V^u_{R} \, u_{\Rhand}
\,,\\
\massbasis{d}_{\Lhand} &= V^d_{L} \,d_{\Lhand}
\,, &
\massbasis{d}_{\Rhand} &= V^d_{ R} \,d_{\Rhand}
\,.
\end{align}
\end{subequations}

Finally, let us have a look at the gauge LQ interactions.
Like in previous sections, we assume that $\nuR$ are heavy due to the inverse seesaw \cite{Mohapatra:1986aw,Perez:2013osa} and therefore their interactions with the $U_1$ leptoquark 
are unimportant for the low-energy phenomenology. 
Interactions of $U_1$ with the other fermions can be rewritten as follows:

\begin{widetext}
\begin{align}
\begin{split}
\Lag &= \frac{g_4}{\sqrt 2} \left(
\dbar{\QL} \, \gamma^\mu \,{}^{4\!}\LL + \dbar{\dR} \, \gamma^\mu \,{}^4\eR
\right) {U_1}_\mu + \hc
\\ 
&=
 \frac{g_4}{\sqrt 2} \Bigg[
	\begflmatrix \dbar{\QLmass} & 0 \enflmatrix
	\gamma^\mu 
	\begflmatrix 
		\VdL & 0 \\ 0 & \unitmatrix 
	\enflmatrix
	\bigg\lflbracket\; \VeL \; \bigg\rflbracket^\dagger 
	\begflmatrix \LLmass \\ L_\Lhand 	\enflmatrix
 +  \begflmatrix\dbar{\dRmass}& 0 \enflmatrix
	\gamma^\mu 
	\begflmatrix \VdR & 0 \\ 0 & \unitmatrix \enflmatrix
	\bigg\lflbracket\; \VeR \; \bigg\rflbracket^\dagger 
	\begflmatrix \eRmass \\  E_\Rhand 	\enflmatrix
\Bigg] {U_1}_\mu + \mathrm{h.c.}
\\
 &= \frac{g_4}{\sqrt 2} \Bigg[
	\begflmatrix\dbar{\QLmass}& 0 \enflmatrix
	\gamma^\mu
	\bigg\lflbracket \; \UL \; \bigg\rflbracket
	\begflmatrix \LLmass \\ L_\Lhand 	\enflmatrix
  +  \begflmatrix\dbar{\hat d_R}& 0 \enflmatrix
	\gamma^\mu
	\bigg\lflbracket \; \UR \; \bigg\rflbracket
	\begflmatrix \eRmass \\ E_\Rhand 	\enflmatrix
\Bigg] {U_1}_\mu  + \mathrm{h.c.}
\label{extendedSU4LQinteractions}
\end{split}
\end{align}
\end{widetext}
The $L_\Lhand$ field on the last line is the heavy $\SUL$ doublet containing $E_\Lhand$ as a component. 
Apparently, the novelty of such extended $\SUC$ models  consists in the fact that the unitary matrices $\ULR$, defined by the last line of \cref{extendedSU4LQinteractions}, are now of dimension $3+\nLRext$. 
Using the block-form notation
\begin{align}
\ULR &= \begflmatrix 
\Unull & \Ulep \\
\Uq & \Uqlep
\enflmatrix_{L,R},
\label{ULRblock}
\end{align}
only the $3\times 3$ submatrices $\Unull_{L,R}$ are relevant for the interactions among the SM fermions.
The larger the numbers $\nLRext$ of extra lepton generations, the more parametric freedom in $\Unull_{L,R}$ is available.
With $\nLext=\nRext=3$, 
one can already choose \emph{any} form of $\frac{g_4}{m_{U_1}} \Unull_{L,R}$, which is all that is relevant for the low-energy phenomenology  at the leading order, cf.~\cref{Wilsons-GLQ}. 

Similar models have already been studied in the literature, usually considering the cases equivalent to $(\nLext,\nRext)= (3,0)$ \cite{Foot:1997pb}, $(0,3)$ \cite{Balaji:2018zna} or $(3,3)$ \cite{Calibbi:2017qbu}. In this work, we focus on the more economical models with $\nLRext < 3$, which are less challenging if one aims to capture all the possible NP signals in the model, but more restrictive if parameters leading to a chosen signal (such as the $b\to s\mu\mu$ anomalies) are searched for.

Note that enlarging the dimension of $\ULR$ is indeed the only practical consequence of extending the theory of QLU from previous sections: we assume that the extra leptons are too heavy to be observed and ignore the details of the scalar sector responsible for the mixing.
A construction of the scalar sector leading to a chosen form of $\ULR$ in similar models can be  found, e.g., in 
Ref.~\cite{DiLuzio:2018zxy}.

Note that although we keep neglecting the $Z'$ in the model, 
it may actually be relevant in some cases. The discussion of this issue is deferred to \cref{app:Zprime}.

\subsection[First signals of gauge LQ in extended SU(4) models]{First signals of gauge leptoquark in extended \texorpdfstring{$\boldsymbol{\SUC}$}{SU(4)C} models}
\label{sec:extended_results}

\newcommand{\nLnRtableIndexColor}{blue!5}
\newcommand{\nLnRtableCornerColor}{blue!10}
\begin{table*} \centering
\setlength{\arrayrulewidth}{0.6pt}
\begin{tabular}{|>{\columncolor{\nLnRtableIndexColor}}C||C|C|C|C|}
\hhline{-||---}\noalign{\vskip.1pt}
\rowcolor{\nLnRtableIndexColor}
\cellcolor{\nLnRtableCornerColor}
\text{Model}
	&	\begin{tabular}{C} 
		\nLext = 0 	\\ \dim \UL = 3 
		\end{tabular}
	&	\begin{tabular}{C}   
		\nLext = 1 	\\ \dim \UL = 4 
		\end{tabular}
	&	\begin{tabular}{C}   
		\nLext = 2 	\\ \dim \UL = 5 
		\end{tabular}
\\ \hhline{=#===}\noalign{\vskip.1pt}
\begin{tabular}{C}   
\nRext = 0 	\\ \dim\UR = 3 
\end{tabular}
	&	\text{see \cref{tab:LimitingBRs}}
	&	\begin{matrix}
	      \BR(B^0 \to ee) \\ 
	      \BR(B_s \to ee) \\
		  (\varepsilon'/\varepsilon)_{K^0}
	     \end{matrix}	
	&	\begin{matrix}
			\BR(B^+\to K^+\mu^+ e^-)\\
			\BR(B^+\to K^+\mu^- e^+)\\
			\varepsilon_{K^0}\\
			\ldots
		\end{matrix}
\\ \hhline{-||---}\noalign{\vskip.1pt}
\begin{array}{c}   
\nRext = 1 	\\ \dim\UR = 4 
\end{array}
	&	\begin{matrix}
	      \BR(B^0 \to ee) \\ 
	      \BR(B_s \to ee) \\
		  (\varepsilon'/\varepsilon)_{K^0}
	     \end{matrix}	
	&	\begin{matrix}
			\varepsilon_{K^0}\\
			\ldots
		\end{matrix}
\\ \hhline{-||--}\noalign{\vskip.1pt}
\begin{array}{c}   
\nRext = 2 	\\ \dim\UR = 5 
\end{array}
	&	\begin{matrix}
			\BR(B^+\to K^+\mu^+ e^-)\\
			\BR(B^+\to K^+\mu^- e^+)\\
			\varepsilon_{K^0}\\
			\RKRKs\\
			\ldots
		\end{matrix}
\\\hhline{-||-}
\end{tabular}
\caption{
Possible \emph{future first  signals} of the gauge LQ  in extended $\SUC$ models featuring $\nLext$ extra lepton doublets and $\nRext$ extra charged-lepton singlets. 
For a given cell, all observables from the cells above and  to the left are implicitly assumed to be included. 
The ellipses indicate that the catalogues in the relevant cell might not be complete.}
\label{tab:obsLimitsExtended}
\end{table*}

We have performed an analysis similar to that described in \cref{sec:method} 
for the extended $\SUC$ models with  $(\nLext, \nRext) = (1,0), (0,1), (2,0), (0,2)$, and $(1,1)$. Some details about the scanning procedure can be found in \cref{app:composite parametrization}.

With growing number of free parameters, more couplings  can be "rotated away" from  $\Unull_{L,R}$ to the other parts of  $\ULR$. New interaction patterns become allowed, 
with lower lower  
limits on $m_{U_1}$.
Naturally, the catalogue of the \emph{first future signals} (the observables which currently constrain $m_{U_1}$ for some form of $\ULR$)
grows with the growing dimensions of these unitary matrices. 
The results are captured in \cref{tab:obsLimitsExtended}.

While a lot of effort has been spent to fully explore the parameter space in the cases $(\nLext,\nRext) = (1,0)$ or $(0,1)$,  the number of parameters for $\nLext + \nRext = 2$ is quite high and we admit that the corresponding lists in \cref{tab:obsLimitsExtended} may not be complete.

\subsection[Addressing neutral current B anomalies]{Addressing neutral current \texorpdfstring{$\boldsymbol B$}{B} anomalies}
\label{sec:GLQ-Banomalies}
During the last decade, several discrepancies in both charged-current and neutral-current $B$-meson decays have been reported \cite{Aaij:2019wad,Aaij:2017vbb,Amhis:2019ckw,Aaij:2014pli,Aaij:2015oid}. 
Plenty New Physics interpretations have been suggested (see, e.g.~\cite{Aebischer:2019mlg,Altmannshofer:2021qrr}), including the $U_1$ leptoquark.
Achieving the setup form \cref{B-anomalies-solution} is meaningless within our restricted model as it requires so low scale of $\SUC$ symmetry breaking that neglecting the other BSM fields would  be inadequate.
Nevertheless, reasonable considerations can be made once only the accommodation of the neutral-current anomalies is sought for. 
These anomalies include the tests of \emph{lepton flavour universality}
\begin{align}
\RKRKs &= \frac{\BR(B\to K^{(*)} \mu\mu)}{\BR(B\to K^{(*)}ee)} 
\label{RK-def} 
\end{align}
with $\RKRKs^\text{exp} < \RKRKs^\SM = 1$ (see \cref{tab:VLQ35benchmark}). 
Further  measurements indicate that the NP effect is in the $b\to s \mu\mu$ channel \cite{Aaij:2014pli,Aaij:2015oid}. 

In order to ascribe this effect to the gauge LQ, the elements $\UL_{s\mu}$ and $\UL_{b\mu}$ need to be non-negligible. 
To avoid the scalar-type operators $\Op_{\ledq}$ involving electrons or muons, which are responsible for the most severe constraints found in \cref{sec:results}, $\nRext = 2$ generations of extra leptonic $\SUL$-singlets are required;
the model with $\dim\left(\UL\right) = 3$ and $\dim\left(\UR\right) = 5$ allows for the following setup:
\begin{align}
\begin{split}
\UL &= \begflmatrix 
 0 & 0 & e^{i\delta_L} \\
 e^{i\delta_1} \cos\gamma & -e^{-i\delta_2} \sin\gamma &  0 \\
 e^{i\delta_2} \sin\gamma & e^{-i\delta_1} \cos\gamma & 0 
\enflmatrix
, \\
  \Unull_{R} &= \begflmatrix
 0 & 0 & 0 \\
 0 & 0 & 0 \\
 0 & 0 & e^{i\delta_R} 
\enflmatrix.
\end{split}
\label{BanomalySolution3L5R}
\end{align}

Note that a similar pattern for $\UL$ has been suggested in Ref.~\cite{Balaji:2018zna} and also in Ref.~\cite{Fornal:2018dqn} within the $SU(4)_{C_L} \times SU(4)_{C_R} \times \SUL\times U(1)_R$ framework where the couplings to the right-handed fermions are suppressed globally.

Adopting \cref{BanomalySolution3L5R}, the maximum likelihood fit is close to the simple case 
\begin{align}
\gamma &=\pi/4
\,, &
\delta_1 = \delta_2 & = \delta_L = \delta_R = 0
\,,&
m_{U_1} &= 22\TeV
,\label{benchmark3L5Rparameters}
\end{align}
which improves the global log-likelihood function of \smelli \cite{Aebischer:2018iyb} by more than 14 units compared to the SM, i.e. $\log(L/L^\SM)\approx 14$.
Such a scenario accommodates well the $\RKRKs$ anomaly and also significantly mitigates the tension in the additional $b\to s\mu\mu$ obser\-vables. 

Using the standard normalization factor $\NN = \frac{4 G_F}{\sqrt{2}} V_{tb} V_{ts}^* \frac{e^2}{16\pi^2}$ for the effective four-fermion operators 
\begin{subequations}
\begin{align}
\Op_{9\sep \qu\qu'\lep\lep'}&=\NN
	\left( \qLprBar \gamma_\mu \qL \right)
	\left( \dbar{\lep'} \gamma^\mu \lep \right)
\,,
\\
\Op_{10\sep \qu\qu'\lep\lep'}&=\NN
	\left(\qLprBar \gamma_\mu \qL \right)
	\left(\dbar{\lep'} \gamma^\mu \gamma_5 \lep\right)
\end{align}
\end{subequations}
in the weak effective theory at the $5\GeV$ scale, \cref{BanomalySolution3L5R,benchmark3L5Rparameters} imply the following contributions of New Physics to the Wilson coefficients:
\begin{subequations}
\begin{align}
  C_{9\sep bs\mu\mu}^\NP 
=+C_{9\sep bs\mu e}^\NP
&= -0.24 
\,,\\
C_{9\sep bsee}^\NP=C_{9\sep bse\mu}^\NP
&= +0.24
\,,\\
 C_{10\sep bs\lep\lep'}^\NP &= -C_{9\sep bs\lep\lep'}^\NP
 \,.
\end{align}
\end{subequations} 
In comparison, the benchmark one-dimensional effective scenario with only $ C_{9\sep bs\mu\mu}^\NP =-C_{10\sep bs\mu\mu}^\NP = -0.53$ \cite{Aebischer:2019mlg} improves log-likelihood to $\log (L/L^\SM) = 18$; the simplified vector LQ setup in \cref{B-anomalies-solution} leads to $\log (L/L^\SM) = 30$ 
as it also accommodates $\RDRDs$.
Note that the discussion in terms of confidence levels would be pointless since these models  differ in number of free parameters.

Predictions for several important observables following from Eqs.~\eqref{BanomalySolution3L5R} and \eqref{benchmark3L5Rparameters} are given in \cref{tab:VLQ35benchmark}.
As outlined in \cref{sec:robustApproach}, the LQ has been integrated out at the tree level and the calculated LFV dipole operators responsible for $\mu\to e \gamma$ arise solely from the one-loop RGE running of the Wilson coefficients. Thus, the predictions for the loop processes should be interpreted with caution.

\begin{table*}\centering
\rowcolors{2}{white}{green!30!blue!8}
\begin{tabular}{L C R >{\hspace{-5px}}l R >{\hspace{-1px}}l}
\hline
\rowcolor{green!30!blue!20}
\text{Observable} 
	& \multicolumn{1}{c}{\begin{tabular}{c}Model\\ prediction\end{tabular}}
	& \multicolumn{2}{c}{Experiment} 
    & \multicolumn{2}{c}{SM prediction}  
\\ \hline
\RK\,[(1.1; 6) \GeV^2]
 & {0.79}
 & {0.85\pm 0.06}\; \;\;& \cite{Aaij:2019wad}
 & 1.00 & \cite{Isidori:2022bzw, Isidori:2020acz,Bordone:2016gaq}
\\
\RKs[(1.1; 6) \GeV^2 ]
 & 0.79
 & 0.68\pm 0.12 \;\;\;&  \cite{Aaij:2017vbb} 
 & 1.00 & \cite{Bordone:2016gaq}
\\
\BR(B_s \to \mu^+\mu^-)
 & 3.2 \times 10^{-9}
 & (3.0 \pm 0.4) \times 10^{-9} \;\;
		& \cite{Zyla:2020zbs}
 & (3.7\pm0.2)\times 10^{-9} & [\texttt{flavio}]
\\ 
\BR(B^+ \to K^+ \mu^+ e^-)
 & 2.1 \times 10^{-9} 
 & <6.4 \times 10^{-9}\; \;& \cite{Aaij:2019nmj}
 & 0&
\\
\BR(B^+ \to K^+ e^+ \mu^-) 
 & 2.1 \times 10^{-9} 
 & <7.0 \times 10^{-9} \;\;& \cite{Aaij:2019nmj}
 & 0&
\\
\BR(\mu\to e\gamma) 
 & \phantom{,}1.9 \times 10^{-13} 
 & < 4.2 \times 10^{-13} & \cite{TheMEG:2016wtm}
 & 0&
\\
\BR(B^0 \to \tau^+ \tau^-) 
 & \phantom{0.}9 \times 10^{-7}
 &  < 1.6 \times 10^{-3}\;\; & \cite{Aaij:2017xqt}
 & 2 \times 10^{-8} & \cite{Bobeth:2013uxa}
\\
\BR(B_s \to e^\pm \tau^\mp) 
 & 6.4 \times 10^{-7}
 & \text{N/A}\;\;\; & \cite{Zyla:2020zbs} 
 & 0&
\\
\BR(B_s \to \mu^\pm \tau^\mp)
 & 6.4 \times 10^{-7} 
 & < 3.4 \times 10^{-5} \;\;& \cite{Aaij:2019okb}
 & 0&
\\
\hline 
\end{tabular}
\caption{Predictions for the benchmark case of \cref{BanomalySolution3L5R,benchmark3L5Rparameters} for several observables with  NP contribution. 
}
\label{tab:VLQ35benchmark}
\end{table*}

In the scenarios with nonzero couplings $\UL_{se}$, $\UL_{be}$, $\UL_{s\mu }$, and $\UL_{b\mu}$
, the strongest bounds arise from  $B^+ \to K^+ \mu^\pm e^\mp$ and from the the LFV loop processes like $\mu\to e \gamma$ (see Ref.~\cite{Crivellin:2017dsk} for a dedicated study).
Generally, the constraints from the latter are quite strong. 
However, in the chiral leptoquark models with unitary interaction matrix, $\mu\to e \gamma$ is suppressed by an analogue of the GIM mechanism. 
As the only non-vanishing element of $\Unull_R$ in \eqref{BanomalySolution3L5R} is essentially irrelevant for $\mu \to e\gamma$, the same applies also to our case. 
Note that Ref.~\cite{Crivellin:2017dsk} did not consider the subleading terms and hence found exactly zero contributions to $\mu\to e\gamma$ for the case $\UL_{se}\UL_{s\mu }^* = -\UL_{be}\UL_{b\mu}^*$.
Ref.~\cite{Fornal:2018dqn} considered the case equivalent to $\UL$ from \eqref{BanomalySolution3L5R} and $\UR = 0$, finding the constraint $m_{U_1}>10\TeV$ based on the \BABAR search \cite{Aubert:2006vb} for $B\to K e\mu$.
The very recent measurement by \LHCb \cite{Aaij:2019nmj} has pushed this limit to 17 TeV for the considered interaction pattern.

Finally, let us note that although the $Z'$ interactions are not lepton-flavour universal, the couplings in the particular case of \cref{BanomalySolution3L5R} are lepton-flavour diagonal and, hence, the $Z'$ does not mediate any flavour violating processes (see \cref{app:LFUV} for more details about lepton flavour).
At the same time, with the mass around $20\TeV$, $Z'$ is also safely hidden to the high-energy searches at LHC. We elaborate on $Z'$ in \cref{app:Zprime}.

To conclude, the interactions of the $\SUC$ gauge leptoquark in a model with two extra weak-isosinglet charged leptons can accommodate the neutral-current $B$-meson anomalies to a large extent.
The suggested scenario can be excluded by future negative searches for $B\to K e\mu$ at \LHCb or \BELLEII.

\section{Conclusions}
\label{sec:conclusion}
We have studied the phenomenology of the gauge leptoquark model with $\SUC$ symmetry of the Pati-Salam type, taking into account the most recent experimental data. 
The catalogue consisting of 11 observables which currently set the border of the excluded part of the parameter space  has been compiled in \cref{tab:LimitingBRs}. 
These observables have a potential to uncover the  gauge LQ signal even with a small improvement of the experimental sensitivity.

For the decays $P^0\to\lep^+ \lep^{\prime-}$ not listed in the catalogue, we have found the future experimental bounds needed in order to further probe the considered model.

Furthermore, we have explored a class of $\SUC$ models with extra heavy vector-like leptons and searched for additional possible future first signals of the gauge LQ. 
We have also found the smallest of these models capable of accommodating the neutral current anomalies in $B$ decays and identified the key future measurement which can exclude such a setup. 

%

\section*{Acknowledgement}
\label{ack}
We acknowledge the support from the Grant agency of the Czech Republic, project no. 20-17490S, from the Grant Agency of Charles University (GAUK) project no. 12481/2019, and from the Charles University Research Center UNCE/SCI/013. 
We would like to express our gratitude to Michal Malinský for his valuable advice and all the support.

\appendix

\section{On group theory of lepton flavour in leptoquark models}
\label{app:LFUV}
For simplicity, let us define the \emph{lepton flavour group in a wider sense} as the $U(3)_\LF$ group acting uniformly by its defining representation on both SM leptonic triplets
\begin{align}
\eRmass  &= \begflmatrix e \\ \mu \\ \tau \enflmatrix_{\!\Rhand}
&&\text{and}&
\LLmass &= \begflmatrix 
\LeptonDoublet_e \\
\LeptonDoublet_\mu \\
\LeptonDoublet_\tau \enflmatrix_{\!\Lhand}.
\end{align}
Note that we have ignored the axial factor of what is usually called the lepton flavour group.
There are three important subgroups of $U(3)_\LF$:
\begin{enumerate}
    \item The \emph{lepton number} group is the Abelian factor emerging in the factorization $U(3)_\LF = SU(3)_\LF \times U(1)_\mathcal{L}$. It acts on $\eRmass$ and $\LLmass$ as multiplication by an overall complex phase.
\item The \emph{lepton flavour group in the strict sense} $U(1)^2_\LF = U(1)_{\Lept_\mu-\Lept_e}\times U(1)_{\Lept_\tau-\Lept_e}\subset SU3)_\LF$ is a group of diagonal special unitary $3\times3$ matrices. In combination with the $\mathcal{L}$ conservation, the $U(1)^2_\mathrm{LF}$ symmetry would imply conservation of the individual lepton family numbers, satisfying $\Lept_e + \Lept_\mu + \Lept_\tau = \Lept$. 
Notice that despite various conventions for what is called the lepton flavour group, the term \emph{lepton flavour violation} (LFV) is being used strictly in relation with $U(1)^2_\mathrm{LF}$.

\item Inspecting non-diagonal parts of the anticipated approximate LF symmetry consists especially in testing the \emph{lepton flavour universality}
(LFU) which can be associated with the group of permutation matrices $(S_3)_\LFU \subset U(3)_\LF$.

\end{enumerate}
Since neither $U(1)^2_\LF$ nor $(S_3)_\LFU$ is a subgroup of the other, LFV
does not necessarily imply LFU violation (LFUV) nor vice versa.

Let us trace the fate of these would-be  symmetries in  leptoquark interactions. For clarity of expression, consider only a single term, say $\dbar \dRmass \,\slashed U_{\!1}\UR\, \eRmass $; the generalization to full-fledged interaction  such as those in \cref{Lag-U1} is straightforward. 
\begin{enumerate}
\item Apparently, the LQ interaction with the leptons and quarks conserves the lepton number $\Lept$ regardless of the form of the interaction matrix $\UR$, provided the $U_1$ leptoquark carries $\Lept = -1$.
\item If two columns of the interaction matrix $\UR$ are zero, then the LQ can be ascribed the corresponding flavour number ($\Lept_e$, $\Lept_\mu$ or $\Lept_\tau$) and there is no LFV. In the case $\UR$ has a single zero column,  only a one-dimensional subgroup of $U(1)^2_\LF$ is a symmetry of the interaction (only the non-interacting flavour remains preserved). If all its columns are non-empty, $U(1)^2_\LF$ is completely explicitly broken. 
\item On the other hand, respecting the $(S_3)_\LFU$ symmetry requires that all three columns of $\UR$ are equal.
Thus, the leptoquark brings new sources of LFUV whenever (at least) two columns of $\UR$ differ. 
\end{enumerate}
These observations hold generally, for any kind of LQ and its interaction matrix. In principle, the form of the interaction matrices may be such that either $U(1)^2_\LF$ or $(S_3)_\LFU$ is an exact  symmetry of the LQ interactions.

However, in the particular case of the gauge LQ in quark-lepton unification, the interaction matrix $\UR$ is a subject of the unitarity conditions: the column normalization rule implies that none of the columns can be empty, the $U(1)^2_\LF$ symmetry is completely broken and the LQ inevitably mediates LFV processes. 
Complementarily, the column orthogonality condition implies violation of $(S_3)_\LFU$. 

In fact, no nontrivial subgroup of $SU(3)_\LF$ can be a symmetry of $\dbar \dRmass \,\slashed U_{\!1}\UR\, \eRmass  $ for any invertable (e.g. unitary)~$\UR$: assuming $X \in U(3)_\LF$ acts as $\eRmass \to X \eRmass $ and $U_1 \to e^{i\varphi(X)} U_1$, the considered interaction
remains intact if and only if $e^{i\varphi(X)} \UR X = \UR$, i.e., if $X$ is a mere phase.

\section[The Z' boson in SU(4) models]{The \texorpdfstring{$\boldsymbol{Z'}$}{Z'} boson in \texorpdfstring{$\boldsymbol{\SUC}$}{SU(4)C} models}
\label{app:Zprime}

\begin{table*}
\centering
{
\rowcolors{1}{orange!22!white}{white}
\begin{tabular}{C||C|C|C|C}
\hline
\GQL
 &\multicolumn{2}{C|}{\SUC }	
 &  U(1)_R 
 & \SUL 
\\ 
\downarrow
& \begin{array}{c}
 		T^{1,\ldots,8}_C \\ 
 		g_3 = g_4 \\
 		G_\mu^{1,\ldots,8}=A_\mu^{1,\ldots,8}
 	\end{array}
 & \begin{array}{c}
 		\BminL = \sqrt\frac{8}{3} T^{15}_C 
 		\\ g_{BL}=\sqrt\frac{3}{8}g_4
 		\\ A_\mu^{15}
 	\end{array}
 & &
\\ 
\GBL
 &  \SUc
 & U(1)_\BminL 
 & U(1)_R & \SUL
\\
\downarrow & 
 &\multicolumn{2}{C|}{
	\begin{array}{c}
		Y=\frac{1}{2}\BminL+R \\ 
		g' = 
		\frac{g_{BL}g_R}{\sqrt{g^2_{BL} + (g_R/2)^2}} 
		= 2 g_{BL} \sin\theta'
		\\
		B_\mu  = \sin\theta' A^{15}_\mu 
					+ \cos\theta' B'_\mu
 	\end{array}}
 &
\\
\GSM & \SUc
 &\multicolumn{2}{C|}{\begin{array}{c} U(1)_Y  \end{array}}
 & \SUL
\\
\downarrow & 
 & \multicolumn{3}{C}{\begin{array}{c}
  Q=T^3_L + Y \\ 
  e=\frac{g g'}{\sqrt{g^2 + g'^2}}= g \sin\theta_W\\
  A_\mu = \cos\theta_W B_\mu-\sin\theta_W W_{\!\mu}^3 
 \end{array}}
\\
\Gvac & \SUc
 & \multicolumn{3}{C}{\begin{array}{c}
  U(1)_Q
 \end{array}}
\\ \hline
\end{tabular}
}

\caption{Scheme of the sequential symmetry breaking in the quark-lepton symmetry scenarios. For each step, the corresponding branching rules, matching equations and gauge bosons which remain massless are specified.}
\label{tab:SU4breaking}
\end{table*}

The features of the $Z'$ boson can be reviewed most naturally when the intermediate gauge symmetry stage
\begin{align}
    \GBL = \SUc \times U(1)_\BminL \times \SUL \times U(1)_R
\end{align}
is considered. The details of the sequential breaking of the $\GQL$ symmetry including this step are summarized in \cref{tab:SU4breaking}.

In the first step of symmetry breaking, 
the $\SUC$ factor is spontaneously broken at some high scale way above the electroweak one, which (unlike for GUTs) can be chosen arbitrarily since our framework unifies the fermions but not the gauge interactions.
The smallest possible first step of the $\SUC$ breaking is 
\begin{align}
\SUC \to \SUc \times U(1)_\BminL\,.
\label{SU4breaking}
\end{align}
The Abelian factor in \cref{SU4breaking} is generated by 
\begin{align}
T^{15}_{C} =\frac{1}{2\sqrt{6}} \begin{pmatrix}
\unitmatrix_{3\times3}&0\\0&-3
\end{pmatrix}.
\label{T15}
\end{align}
Quite commonly, its multiple
\begin{align}
\BminL = \sqrt{\frac{8}{3}} T^{15}_C =
\mathrm{diag}\left(\sfrac 1 3,\sfrac 1 3,\sfrac 1 3,-1\right)
\label{defOfBminLgenerator}
\end{align}
is being used instead, which is compensated by redefinition of the gauge coupling: $g_{BL} = \sqrt{3/8} \,g_4$. 

The name of the $\BminL$ generator is motivated by 
its action on the unified fermionic representations in \cref{fermionsIn4plets}. 
However, one must keep in mind that action of this symmetry generator in \cref{defOfBminLgenerator} does not necessarily coincide with the difference between the baryon number $\Bary$ and lepton number $\Lept$ for other fields in the model.
For example, in the Minimal Quark-Lepton Symmetry Model \cite{Smirnov:1995jq}, both $\Bary$ and $\Lept$ are perturbatively conserved to all orders while $T^{15}_C$ is spontaneously broken. 
Next, the extended models studied in \cref{sec:extended} contain leptonic fields ${}^1\LeptonDoublet$ and ${}^1 e$ which transform trivially under $\SUC$ and hence also under $U(1)_\BminL$.
To emphasise the distinction between $\Bary-\Lept$ and the gauge symmetry generator \eqref{defOfBminLgenerator}, we shall  keep the square brackets around the latter in order to indicate that $\BminL$ is an indivisible symbol.




The 19 gauge fields of the model can be cast as follows:
\begin{subequations}
\begin{align}
\SUC: &\qquad 
\mathcal A_\mu = 
\begin{pmatrix}
	   G_\mu +\frac{1}{2\sqrt{6}}A^{15}_\mu & {U_1}_\mu/\sqrt{2} 
	\\ {U_1}_\mu^\dagger/\sqrt{2} & -\frac{3}{2\sqrt{6}}A^{15}_\mu
\end{pmatrix} \label{SU4gaugeFields}
\\
\SUL:&\qquad
\mathcal W_\mu = \frac{1}{2}
\begin{bmatrix}
W^3_\mu & \sqrt{2} \, W^+_\mu \\ \sqrt{2} \, W^-_\mu & - W^3_\mu
\end{bmatrix}
\\
U(1)_R:&\qquad
B'_\mu\end{align}
\end{subequations}
In \cref{SU4gaugeFields}, the $(3\!+\!1)\times (3\!+\!1)$ block notation has been used. Together with the gluons $G$ and charged intermediate vector bosons $W^\pm$, one can easily identify the vector leptoquark ${U_1}$. Furthermore, the three electrically neutral fields $A^{15}, B'$, and $W^3$ mix into the photon, the $Z$ boson, and to $Z'$.

The symmetry breaking \eqref{SU4breaking} gives mass only to the gauge leptoquark; the $Z'$ boson acquires mass no sooner than during the second step,
\begin{align}
    U(1)_\BminL \times U(1)_R \to U(1)_Y\,.
    \label{BminL_breaking}
\end{align}
Thus, while the precise ratio of $m_{U_1}/m_{Z'}$ depends on the scalar sector of the model, $Z'$ can never be much heavier than $U_1$. 

\begin{widetext}
The rotation of the electrically neutral gauge fields to the mass basis can be written as
\begin{align}
\begin{pmatrix} 
   A^{15}_\mu \\ B'_\mu \\ W^3_{\!\mu}
\end{pmatrix}
=
\begin{pmatrix}
  \cos\theta'  &  \sin\theta' & 0 \\
 -\sin\theta'  &  \cos\theta' & 0 \\
 0  &  0  &  1
\end{pmatrix}
\begin{pmatrix}
  \ZOprime_\mu \\ B_\mu \\ W^3_{\!\mu}
\end{pmatrix}
=
\begin{pmatrix}
  \cos{\theta'}  &  \sin{\theta'} & 0 \\
 -\sin{\theta'}  &  \cos{\theta'} & 0 \\
  0   &  0  &   1
\end{pmatrix}
\begin{pmatrix}
  1  &  0  &  0 \\
  0 & \cos\theta_W  &  \sin \theta_W \\
  0 & -\sin \theta_W  &  \cos \theta_W
\end{pmatrix}
\begin{pmatrix}
   \cos \theta_{m}  &  \sin \theta_{m} & 0 \\
 -\sin \theta_{m}  &  \cos \theta_{m} & 0 \\
  0   &  0  &   1
\end{pmatrix}
\begin{pmatrix}
  Z'_\mu \\ Z_\mu \\ A_\mu
\end{pmatrix},
\label{ZZgammaRotations}
\end{align}
\end{widetext}
where $\tan \theta' = g_R / (2 g_{BL})$ at the relevant scale, and $\theta_W$ is the  weak mixing angle  (see  \cref{tab:SU4breaking}). 
The angle $\theta_m$ is very small when the symmetry breaking \eqref{BminL_breaking} occurs way above the electroweak energy scale \cite{Smirnov:1995jq}.
Hence, in the limit $m_{Z'} / m_{Z} \to \infty$,
 the $Z'$ boson is given by 
\begin{align}
Z'_\mu = A^{15}_\mu \cos \theta' - B'_\mu \sin \theta'
\,.
\end{align}
and the $Z'$ coupling can be obtained by rewriting the relevant terms in the covariant derivative using relations from \cref{tab:SU4breaking},
\begin{align}
\begin{split}
g_{BL} &\BminL A^{15}_\mu + g_R R B'_\mu
\\&=
g' Y B_\mu + \frac{g_{BL}}{\cos\theta'}
\left( \BminL - 2Y\sin^2\theta'\right) Z'_\mu
\,,
\end{split}
\label{Zprime-coupling}
\end{align}
which is an analogue to the SM case
\begin{align}
\begin{split}
g' Y B_\mu &+ g T^3_L W^3_{\!\mu}
\\&=
e \,Q A_\mu  + \frac{g}{\cos\theta_w} 
\left( T^3_L - Q \sin^2\theta_w \right) Z_\mu
\,.
\end{split}
\label{Z+gamma-interactions}
\end{align}

In the models of QLU, where all the fermions arise from $\SUC$ quadruplets, the $Z'$ interactions with both quarks and leptons are flavour-diagonal and universal, i.e., they respect the entire $U(3)_\LF$ symmetry. 
The coupling strength is governed by \cref{Zprime-coupling}.
With the $\SUC$ breaking scale around 100~TeV or higher, the resulting flavour-conserving 4-fermion operators are safely negligible in the simplest situations without the optional symmetry-breaking step of \cref{SU4breaking}.
On the other hand, the role of $Z'$ in the extended models might  be much more important since the mass limits are generally lower and its interactions with the leptons do not necessarily conserve flavour. 

Lepton-flavour conserving effective semileptonic interactions mediated by  $Z'$ could interfere with the SM amplitudes in the $q\dbar q  \xrightarrow{Z^*\!,\, Z'{}^*} \lep^+  \lep^-$ production in the $s \gg m_Z^2$ kinematic region. 
NP contributions to these processes are constrained by the high-$p_T$ dilepton spectra measurements by \ATLAS and \CMS, leading to limits around $m_{Z'} > 5\TeV$ (depending on the $Z'$ coupling assumed) \cite{Aaboud:2017buh,Sirunyan:2018exx}.
As noted in Ref.~\cite{Baker:2019sli}, these limits also indirectly constrain the mass of the gauge LQ. This bound is important in models accommodating the anomalous value of $\RD$ which require $m_{U_1} \sim 2\TeV$.

Ref.~\cite{Baker:2019sli} further states that "\emph{the couplings of the $Z'$ to SM fermions are necessarily flavour universal}" and "\emph{proportional to the identity matrix in flavour space}"	even in the models with extra fermions because the relevant charged lepton mixing \emph{"necessarily involve states with the same $B-L$ charge"}.
This is, however, a misconception arising from not-distinguishing between the gauge symmetry generator $\BminL$ and the difference of the accidental global symmetries $\Bary - \Lept$. 
All the fermionic fields ${}^4\LL, {}^4\eR, {}^1\LLR, {}^1\eLR$ are fully justified to be called \emph{leptons} and carry the lepton number $\Lept$, which is conserved by the gauge interactions. 
On the other hand, only the fields ${}^4\LL$ and ${}^4\eL$, which stem from $\SUC$ quadruplets, are also charged with respect to~$\BminL$, the diagonal generator of the $\SUC$ group. 
\begin{widetext}As a consequence of this, rotating the the left-handed $\left(\BminL - 2Y\sin^2\theta'\right)$ lepton currents into the mass basis [see \cref{extendedSU4LQinteractions}] yields
\begin{subequations}
\begin{align}
\begflmatrix
\dbar{{}^4\LL} & \dbar{{}^1\LL} 
\enflmatrix
\begflmatrix
-1+\sin^2\theta'  & 0 \\ 0 & \sin^2\theta'
\enflmatrix
\gamma^\mu
\begflmatrix
{}^4\LL \\ {}^1\LL 
\enflmatrix
&=
\begflmatrix
\dbar{\LLmass} & \dbar{L_\Lhand}
\enflmatrix
\VeL
\begflmatrix
-1+\sin^2\theta'  & 0 \\ 0 & \sin^2\theta'
\enflmatrix
\VeL ^\dagger\,
\gamma^\mu
\begflmatrix
\LLmass \\ L_L 
\enflmatrix
\end{align}
and similarly for the right-handed currents:
\begin{align}
\begflmatrix
\dbar{{}^4\eR} & \dbar{{}^1\eR}
\enflmatrix
\begflmatrix
-1+2\sin^2\theta'  & 0 \\ 0 & 2\sin^2\theta'
\enflmatrix
\gamma^\mu
\begflmatrix
{}^4\eR \\ {}^1\eR 
\enflmatrix
&=
\begflmatrix
\dbar{\eRmass} & \dbar{E_\Rhand}
\enflmatrix
\VeR
\begflmatrix
-1+2\sin^2\theta'  & 0 \\ 0 & 2\sin^2\theta'
\enflmatrix
\VeR ^\dagger\,
\gamma^\mu
\begflmatrix
\eRmass \\ E_R
\enflmatrix.
\end{align}
\end{subequations}
Finally, using the implicit definition of $\ULR$ in \cref{extendedSU4LQinteractions} and the block notation of \cref{ULRblock}, one arrives to the following formula for the $Z'$ couplings with the SM fermions:
\begin{align}
\begin{split}
\Lag^{Z'\lep \lep} = \frac{g_{BL}}{\cos\theta'}
&\Big[
  \;\dbar\LLmass_i
	\left(
		 s'^2\unitmatrix   
		-(\Unull_L)^\dagger \Unull_L
	\right)
	\gamma^\mu \LLmass^i
+\dbar\eRmass
	\left(
		2 s'{}^2 \unitmatrix 
		- (\Unull_R)^\dagger \Unull_R
	\right)
	\gamma^\mu \eRmass	
\\&
  +\frac{1+s'^2}{3}\, \dbar \QLmass_i \gamma^\mu \QLmass^i 
  +\frac{1-4s'^2}3 \,\dbar\uRmass \gamma^\mu \uRmass
  + \frac{1+2s'^2}3 \,\dbar\dRmass \gamma^\mu \dRmass
\Big] Z'_\mu\,,
\end{split}
\end{align}
\end{widetext}
where $s'{}^2 \equiv \sin^2\theta'$ amounts to $0.08$ at the $2\TeV$ scale or to $s'{}^2 \simeq 0.12$ in the $200\TeV$ ballpark (assuming SM-like gauge coupling running up to $m_{Z'}$).
Thus, the $Z'$ interactions with leptons in the extended $\SUC$ models are not necessarily flavour-universal and, in general, the diagonal couplings could actually be strongly suppressed.

As a consequence, the limits on $m_{Z'}$ from the high-energy dilepton spectra may be considerably weakened for certain patterns of $\Unull_{L,R}$.
The simplified reasoning of Ref.~\cite{Baker:2019sli} mentioned above 
has been used as a no-go argument for abandoning the models with the $\GQL$ gauge group and focusing on $G_{4321}$-based models instead when attempting to accommodate $\RDRDs$. 
In this respect, we note that achieving the form of $\Unull_{L,R}$ from \cref{B-anomalies-solution} in the  framework of extended $\GQL$ models would imply that the $Z'$ couplings to the $e$ and $\mu$ leptons are suppressed. 
Since the $Z'{}^* \to \tau^+\tau^-$ channel is experimentally less constrained \cite{Aaboud:2017sjh}, a valid no-go argument needs to be more subtle.
Nevertheless, the scenarios with the $\SUC$-breaking scale as low as 2 TeV require full model specification since the effects of the new scalar and fermionic degrees of freedom would be important. This is far beyond the scope of this paper.


In any case, this study is focusing on the extended $\SUC$ models with $\nLext + \nRext \leq 2$. 
Such frameworks can not accommodate the $\RDRDs$ anomalies even if the $Z'$ is completely ignored due to the residual constraints on the leptoquark interaction matrices $\Unull_{L,R}$ from the unitarity of $\ULR$.


During scanning of the parameter space of these models, we have not encountered a parameter point allowing for $m_{U_1}$ smaller than 18~TeV. 
Since the models allow for a similarly heavy $Z'$, the constraints from this field are not severe: unlike the gauge LQ, $Z'$ does not contribute to the scalar-type 2-quark-2-lepton operators $\Op_{\ledq}$ but only to the Wilson coefficients multiplying the vector-type ones ($\Op_{ed},\, \Op_{\LeptonDoublet\QDoubletIndex}$) and further to those of flavour-conserving 4-lepton or 4-quark operators, all of which are experimentally less restricted.

In this analysis, the $Z'$ contributions  to the Wilson coefficients are not calculated. 
Including them  could be a part of a future  study focusing on the extended $\SUC$ models.

\section{Optimizing the scanning procedure}
\label{app:composite parametrization}

The experimental data collected over the last decades provided rather stringent constraints on mass of the considered leptoquark. Some of the most restraining processes are the  decays of $\KL\to l^+ l'^-$ and the $\mu\to e$ conversion on gold nuclei, see \cref{tab:LimitingBRs}. Here we identify  areas in the parameter space in which these decays are suppressed, 
and thus allow for lighter leptoquark.

\subsection{\texorpdfstring{Avoiding $\KL\to l^+ l'^-$}{Avoiding KL -> l+l-}
}
The scanning procedure mentioned in \cref{sec:scanning} is optimized when we restrict the parameter space to a subspace in which
\begin{align}
    \BRV(\KL\to l l') = 0\,.
\label{KL=0}
\end{align}

Schematically, the $\UL$ and $\UR$ matrices read
\begin{equation}
\left(
\begin{array}{ccc|c}
\U_{d e} &\U_{d\mu} & \U_{d\tau}& \\
\U_{s e} & \U_{s\mu }& \U_{s\tau}&\Ulep\\
\U_{b e }& \U_{b\mu }&\U_{ b\tau} &\\
\hline
\rule{0pt}{0.9\normalbaselineskip}
&\Uq&&\Uqlep
\end{array}
\right)
_{L,R}
\end{equation}
where $u$, $d$, $s$ are quarks, $e$, $\mu$, $\tau$ are leptons and each element represents the strength of interaction of these two fermions with the leptoquark. The block matrices $\Ulep$,  $\Uq$, and $\Uqlep$ are present only in the extended models studied in \cref{sec:extended}. As follows from Eqs. \eqref{brv} -- \eqref{KS,KL}, the $\BR$'s of the leptonic $\KL$ decays are proportional to
\begin{subequations}\label{all beta}
\begin{align}
\begin{split}
\label{beta emu}
\beta_{\KL,e\mu}^2=
\beta_{\KL,\mu e}^2
=&\;
\tfrac 12 \left|\UL_{de}\UR_{s\mu}^*+\UL_{se}\UR_{d\mu}^* \right|^2
\\+&\; \tfrac 12  \left|\UL_{d\mu}\UR_{se}^*+\UL_{s\mu}\UR_{de}^* \right|^2
\end{split}
\\
\beta_{\KL,ee}^2
=&
\left|\UL_{de}\UR_{se}^* +\UL_{se}\UR_{de}^*\right|^2,\\
\beta_{\KL,\mu\mu}^2=&
\left|\UL_{d\mu}\UR_{s\mu}^*+\UL_{s\mu}\UR_{d\mu}^*\right|^2.
\end{align}
\end{subequations}
All these $\beta$'s vanish if and only if
\begin{align}
\label{all beta matrix form}
\begin{pmatrix}
\UL_{d e} & \UL_{s e} \\
\UL_{d\mu} & \UL_{s\mu}
\end{pmatrix}
\begin{pmatrix}
\UR_{s\mu} & \UR_{d \mu} \\
\UR_{s e} & \UR_{d e}
\end{pmatrix}^{\!*}
=\begin{pmatrix}
0 & 0 \\ 0 & 0
\end{pmatrix}.
\end{align}
. 


When we think of $\UL_{ql}$ as fixed numbers and  $\UR_{ql}$ as the unknowns, the necessary condition for a nontrivial solution to exist shrinks to 
\begin{align}\label{det rule}
\begin{vmatrix}
\UL_{de}&\UL_{d\mu}\\
\UL_{se}&\UL_{s\mu}
\end{vmatrix}=0,
\end{align}
where $\left|\U\right|$ stands for determinant.
On the other hand, we can treat $\UR_{ql}$ as fixed numbers and $\UL_{ql}$ as variables, which leads to analogous result for  the $\UR$ matrix. 
Hence, the determinants of the top left $2\times2$ submatrices of both $\UL$ and $\UR$ has to be equal to zero, regardless of the dimensionality of these matrices. 

Now we use a simplified rule of Laplace expansion in multiple rows (as derived by Laplace in 1772, more on this e.g. \cite{prasolov1994problems})
\begin{equation}\label{laplace}
\left|M^{-1}\right|_{IJ}=\pm\frac{\;\left|M\right|_{I'J'}}{\;\left|M\right|_{\phantom{I'J'}}},
\end{equation}
where  $\left|M\right|_{IJ}$ is the $IJ$-minor, i.e., the determinant of the submatrix obtained from $M$ by deleting rows and columns from  sets $I,J
\subset D=\{1,\dots,\mathrm{dim}(M)\}$
. The set $I^\prime$ ($J^\prime$) is the complement of $I$ ($J$) in $D$, so that every row and every column index appears exactly once in \cref{laplace}. If $M$ is a unitary matrix, its determinant is just a complex phase, and we can further simplify \cref{laplace} to
\begin{equation}\label{laplace for unitary}
\textit{phase} \times \left|M\right|_{IJ}=\left|M\right|_{I'J'}.
\end{equation}
In other words, the determinant of any submatrix of a unitary matrix is equal in magnitude to the determinant of the complementary submatrix. 
Applying this observation to the $\UL$ and $\UR$ matrices of dimension $3$, \cref{det rule} leads to
\begin{equation}\label{eq:bottom right element}
\UL_{b\tau}=\UR_{b\tau}=0
\end{equation}
with an implication $\BRV(\Bds\to\tau\tau)=0$ (which has been derived also in Ref.~\cite{Smirnov:2018ske} by directly solving Eqs.~\eqref{all beta} in a specific parametrization).
Note also that the ”solutions” to the $B$~anomalies which leave out the unitarity constrains such as \cref{B-anomalies-solution}, usually require that the $\ULR_{b\tau}$ elements are the largest ones in order to address also $\RDRDs$. 
Therefore, the anomalous value of $\RDRDs$ can not be accounted to the Pati-Salam-type leptoquark.
%
%

Fulfilling the rather simple condition \eqref{det rule} can be tough for $\UL$, $\UR$ of higher dimensions. 
To this end, we introduce the composite parametrization of $U(n)$ matrices \cite{Spengler_2010,Spengler_2012}, which turns out to be particularly convenient in this respect. 
Its $n^2$ parameters $\lambda_{ij}$ consist of $\tfrac12 n(n-1)$ angles ($i<j$) and $\tfrac12 n(n+1)$ phases ($i\geq j$).
A $3\times3$ matrix in this parametrization reads
\begin{widetext}
\begin{align}
\begflmatrix
 c_{12} c_{13} e^{i \lambda _{11}} 
 & \left(c_{23} s_{12}-c_{12} s_{13} s_{23} e^{i \lambda _{32}}\right) e^{i \lambda _{22}} 
 & \left(s_{12} s_{23}+c_{12} c_{23} s_{13} e^{i \lambda _{32}}\right)  e^{i \lambda_{33}}
\\
 -c_{13} s_{12} e^{i \lambda _{11}+i \lambda _{21}}
& \left(c_{12} c_{23} \!+\! 
  s_{12} s_{13} s_{23} e^{i\lambda_{32}}\right)
  \!e^{i \lambda _{22}+ i \lambda _{21}}
& \left(c_{12} s_{23} -c_{23} s_{12} s_{13} e^{i \lambda _{32}}\right) \!e^{i \lambda_{33}+i\lambda_{21}} 
\\
 -s_{13} e^{i \lambda _{11}+i \lambda _{31}}
 &- c_{13} s_{23} e^{i \lambda _{22}+i \lambda _{31}+i \lambda _{32}} 
 & c_{13} c_{23} e^{i \lambda _{31}+i \lambda _{32}+i \lambda _{33}} 
\enflmatrix
\label{compositePar}
\end{align}
\end{widetext}
where $c_{ij} = \cos\lambda_{ij}$ and $s_{ij}=\sin\lambda_{ij}$.
For higher dimensions we refer to the original literature \cite{Spengler_2010,Spengler_2012} or to the publicly available implementation in \WM \cite{compositeWWW}.
The slice of the parameter space used for \cref{fig:picr} is obtained by fixing these parameters according to \cref{tab:labmdas_picr}.

\begin{table}\centering
{\renewcommand{\arraystretch}{1.15}
\rowcolors{0}{gray!10!}{white}
\begin{tabular}{CCCCCCCCCCCCCCCCCC}
\hline
 \lambda _{11 L} & \lambda _{12 L} & \lambda _{13 L} & \lambda _{21 L} & \lambda _{22 L} & \lambda _{23 L} & \lambda _{31 L} & \lambda _{32 L} & \lambda _{33 L} 
 \\[2pt] \noalign{\vskip 3pt}
 \dfrac{\pi }{7} & \dfrac{\pi }{16} & \dfrac{\pi }{7} & -\dfrac{\pi }{7} & \dfrac{\pi }{7} &(0,2\pi) & \dfrac{\pi }{7} & \dfrac{\pi }{7} & \dfrac{\pi }{7} 
 \\[8pt]\hline 
 \lambda _{11 R} & \lambda _{12 R} & \lambda _{13 R} & \lambda _{21 R} & \lambda _{22 R} & \lambda _{23 R} & \lambda _{31 R} & \lambda _{32 R} & \lambda _{33 R} 
\\[2pt] \noalign{\vskip 3pt}
 \dfrac{\pi }{7} & -\dfrac{\pi }{16} & \dfrac{47 \pi }{96} & \dfrac{\pi }{7} & \dfrac{\pi }{7} & \dfrac{\pi }{2} & \dfrac{\pi }{7} & \dfrac{\pi }{7} & \dfrac{\pi }{7} 
 \\[8pt] \hline
\end{tabular}
}
\caption{Angles and phases defining the slice of the parameter space used in \cref{fig:picr}.}
\label{tab:labmdas_picr}
\end{table}

The beautiful advantage of the composite parametrization is that the subspace obeying \cref{KL=0} can be obtained by fixing the same parameters for \emph{any dimension $\geq 3$ of $\UL$ and $\UR$}. 
The necessary condition~\eqref{det rule} for existence of the solution  is fulfilled by  
\begin{equation}
\label{nes condition}
\lambda_{23L}=\lambda_{23R}=\frac{\pi}{2}.
\end{equation}
With this in hand, it can be 
shown that setting
\begin{subequations}
\begin{align}
\lambda_{12L}=&-\lambda_{12R},\\
\lambda_{21L}=&-\lambda_{21R}
\end{align}
\label{general solution}
\end{subequations}
solves Eqs.~\eqref{all beta matrix form} entirely.
%
Naïvely, other solutions 
can be found but they fall outside the proper domain of the $\lambda$'s. 
An equivalent solution to \eqref{all beta} was found in Ref.~\cite{Smirnov:2018ske} for $\text{dim }\UL = \text{dim }\UR = 3$ within a different parametrization.


\subsection{\texorpdfstring{Avoiding $\CRmueAu$}{Avoiding CR(mu->e, Au)}
}
Leaving the $\KL$ decays for a while, we now focus on another very important constraint stemming from the limits on $\mu\to e$ conversion on gold nuclei, $\CRmueAu < 7\times 10^{-13}$ \cite{Bertl:2006up}. In the same manner, we enforce 
\begin{equation}
    \CRmueAu =0.
\end{equation}

A  leptoquark with $Q=+\sfrac23$ mediates this process at the tree level by an interaction with the $d$ quarks and the sea $s$ quarks in the nucleons. 
The calculation in \flavio is based on Ref.~\cite{Kitano:2002mt}.
The scalar-type effective vertices, $(\dbar {d_\Rhand} d_\Lhand)( \dbar{e_\Lhand} \mu_\Rhand)$ and $(\dbar {d_\Rhand} d_\Lhand )( \dbar{\mu_\Lhand} e_\Rhand)$, are predicted to engage in this process even more efficiently than the vector-type ones. 
Thus, to avoid these constraints when searching for limits from other interesting processes, the following condition must be approximately fulfilled: 
\begin{align}
\label{avoidmue}
|\UL_{de}\UR_{d\mu}^*|^2 +  |\UR_{de}\UL_{d\mu}^*|^2 = 0.
\end{align}

It can be shown that any $\ULR$ pair obeying \cref{avoidmue} together with the set of Eqs.~\eqref{all beta}  
must necessarily have some of the  elements from the upper left $2\times2$ submatrix equal to zero. The possible patterns for $\ULR$ are
\begin{subequations}
\label{rightOut}
\begin{align}
  \UL = \left(
  \begin{array}{cc|c}
  \XXX  &0 &  \\
  \XXX & 0 &  \\
  \hline
  &&
  \end{array}
  \right)
  ,\;\,&
 \UR = \left(
 \begin{array}{cc|c}
  \XXX  &0 &  \\
  \XXX & 0 &  \\
  \hline
  &&
  \end{array}
  \right);
 \label{musOut}
\\
  \UL = \left(
 \begin{array}{cc|c}
   0 & \XXX &  \\
   0&  \XXX &  \\
  \hline
  &&
  \end{array}
  \right) ,\;\,&
 \UR = \left(
 \begin{array}{cc|c}
  0 & \XXX &  \\
   0&  \XXX &  \\
  \hline
  &&
  \end{array}
  \right);
 \label{electronsOut}
\\
   \UL = \left(
 \begin{array}{cc|c}
  0 &0 &  \\
  \XXX & \XXX &  \\
  \hline
  &&
  \end{array}
  \right) ,\;\,&
 \UR = \left(
 \begin{array}{cc|c}
  0 & 0&  \\
  \XXX & \XXX &  \\
  \hline
  &&
  \end{array}
  \right);
\label{d-quarksOut}
\\
   \UL = \left(
 \begin{array}{cc|c}
  0 &0 &  \\
  0 & 0 &  \\
  \hline
  &&
  \end{array}
  \right) ,\;\,&
 \UR = \left(
 \begin{array}{cc|c}
  \XXX & \XXX&  \\
  \XXX & \XXX &  \\
  \hline
  &&
  \end{array}
  \right);
\\
 \UL = \left(
 \begin{array}{cc|c}
  \XXX & \XXX&  \\
  \XXX & \XXX &  \\
    \hline
  &&
  \end{array}
  \right) ,\;\,&
 \UR = \left(
 \begin{array}{cc|c}
  0 &0 &  \\
  0 & 0 &  \\
  \hline
  &&
  \end{array}
  \right),
\end{align}
\end{subequations}
where $\XXX$ denotes an unfixed value. The last two cases are available only when $\UL$ or $\UR$ has dimension $n\geq 4$, respectively.  



Finding the unitary parametrization fulfilling both \cref{avoidmue,all beta} is  straightforward though somewhat tedious as 
the solution has to be found for each dimension of $\UL$ and $\UR$ separately. 

Notable but order-of-magnitude smaller contributions to the coherent $\mu\to e$ conversion  still arise from vector-type operators (triggered by $\UL_{de}\UL_{d\mu}^*$ and $\UR_{de}\UR_{d\mu}^*$) and well as the muon conversion on the sea $s$-quarks in the nucleons (such amplitudes are proportional to $\UL_{se}\UR_{s\mu}^*$ or $\UL_{se}\UR_{s\mu} ^*$).


\bibliography{biblijografije}

\end{document}